\documentclass[aps,twocolumn,groupedaddress,amsmath,ams-fonts,showpacs]{revtex4}
\usepackage[dvipdfmx]{graphicx}
\usepackage{dcolumn}
\usepackage{bm}
\usepackage{here}
\usepackage{relsize}
\usepackage{amssymb}
\usepackage{txfonts}
\usepackage{braket}
\usepackage{color}

\usepackage[pdftex]{hyperref}   
\def\[#1\]{\begin{align}#1\end{align}}

\begin{document}

\title{Supercurrent-induced Skyrmion dynamics and Tunable Weyl points in Chiral Magnet with Superconductivity}
\author{Rina Takashima$^{1}$}
\author{Satoshi Fujimoto$^{2}$}
\affiliation{$^1$Department of Physics, Kyoto University, Kyoto 606-8502, Japan}
\affiliation{$^2$Department of Materials Engineering Science, Osaka University, Toyonaka 560-8531, Japan}
\date{\today}
\begin{abstract}
Recent studies show superconductivity provides new perspectives on spintronics. We study a heterostructure composed of an $s$-wave superconductor and a cubic chiral magnet that stabilizes a topological spin texture, a skyrmion. We propose a supercurrent-induced spin torque, which originates from the spin-orbit coupling, and we show that the spin torque can drive a skyrmion in an efficient way that reduces  Joule heating. 
We also study the band structure of Bogoliubov quasiparticles and show the existence of Weyl points, whose positions can be controlled by the magnetization. 
This results in an effective magnetic field acting on the Weyl quasiparticles in the presence spin textures. Furthermore, the tilt of the Weyl cones can also be tuned by the strength of the spin-orbit coupling, and we propose a possible realization of type-II Weyl points.    
\end{abstract}

\pacs{}
\maketitle

\section{introduction}
Topological structures in states of matter exhibit interesting features beyond topological stability. 
Topological insulators have surface Dirac cones, and show the magnetoelectric polarizability~\cite{Hasan2010, Qi2011}.  
Weyl points in a band structure are predicted to produce a variety of phenomena such as Fermi-arc surface states, chiral anomalies, and unusual anomalous Hall effects~\cite{Volovik2003, Murakami2007,Wan2011, Xu2011,Yang2011b,Burkov2011,Halasz2012, Witczak-Krempa2012,Hosur2012,Zyuzin2012,Aji2012,Son2012,Liu2013}. In a spin texture, of particular recent interest is a magnetic skyrmion \cite{Rossler2006, Muhlbauer2009,Yu2010a,Munzer2010,Yu2011,Nagaosa2013}, which is defined by an integer topological number. Skyrmions are experimentally observed in some chiral magnets such as the cubic $B20$ compounds MnSi and Fe$_{0.5}$Co$_{0.5}$Si. Their topological structure gives rise to a Magnus force that acts on a skyrmion like a Lorentz force~\cite{Stone1996}. When electrons coupled to skyrmions, the so-called emergent electromagnetic field acts on electrons \cite{Volovik1987,Nagaosa2013}. 

Skyrmions have received attentions also as a promising candidate for a future information career. The Magnus force prevents a skyrmion from being pinned by impurities or lattice defects, and the threshold current density to drive its motion is quite low compared to that of a domain wall \cite{Jonietz2010, Yu2012a,Iwasaki2013, Fert2013, Lin2013, Schutte2014c}. 
Such current-induced magnetization dynamics has been studied intensively in the field of spintronics. A spin transfer torque~\cite{Slonczewski1996, Slonczewski1999,Berger1996, Berger1999} is 
one type of current-induced spin torque, which requires noncollinear spin textures. Spin-orbit (SO) coupling realizes another type of current-induced torque, the so-called SO torque~\cite{Chernyshov2008,Manchon2008,Garate2009,Miron2010,Liu2012}, which can also drive domain walls~\cite{Khvalkovskiy2013} or skyrmions via applied currents~\cite{Hals2013,Hals2014}.  

Recently, more and more studies have shown that superconductivity provides new perspectives on spintronics, {such as utilizing a supercurrent for spin dynamics}\cite{Bergeret2001, Waintal2002, Bergeret2005, Keizer2006,Eschrig2008, Zhao2008,Braude2008, Konschelle2009,Eschrig2011, Linder2011, Sacramento2011a,Linder2012, Bergeret2013, Bergeret2014,  Kulagina2014,Halterman2015a,Linder2015,Eschrig2015a,Jacobsen2015,
Yokoyama2015a, Hals2016}.
{Supercurrents can realize efficient ways to control magnetization reducing Joule heating.  With superconductivity, the Joule heating would be neglected since normal currents do not flow, while in a conventional setup with a normal metal, the Joule heating occurs depending on the applied current density. For example, the threshold current density for skyrmion manipulation is $\sim 10^6$ A/m$^2$, but higher motion of skyrmions requires much larger current density~\cite{Jonietz2010,Iwasaki2013}}.
From a theoretical point of view, systems with superconductivity and magnetization are also interesting in the band topology of Bogoliubov quasiparticles.  
With Rashba SO coupling~\cite{Fujimoto2008,Sato2009,Sau2010,Pershoguba2015a, Bjornson2015} or spin textures~\cite{Nakosai2013,Klinovaja2013, Chen2015a,Li2016}, the states of quasiparticles can have nontrivial topology. 
{ Such topological structures of quasiparticle have been extensively studied in the A phase of $^3$He, which has Weyl points in quasiparticle spectrum~\cite{Volovik2003}. The effect of transport properties of Weyl particles on vortex dynamics in superfluid has been discussed~\cite{Bevan1997, Volovik1998}. It would be interesting to propose the application of topological transport phenomena of quasiparticle to spintronics, and systems with superconductivity and magnetism are expected to offer a variety of novel spintronics phenomena.} 

In this paper, we study supercurrent-induced skyrmion dynamics in a heterostructure composed of an $s$-wave superconductor and a cubic chiral magnet (e.g., MnSi and Fe$_{0.5}$Co$_{0.5}$Si). To this end we focus on the  SO coupling in chiral magnets which can be written as  
$
 H_{so}=\alpha_{so} \bm k \cdot \bm \sigma \label{soc}
$ around the $\Gamma $ point, with $\bm \sigma$ being the Pauli matrices acting on the spin space.  
This SO coupling plays a crucial role in chiral magnets, since it leads to the Dzyaloshinskii-Moriya interaction, which is the origin of the noncollinear magnetic orders such as helical spin orders and skyrmion crystal phases. With the SO coupling, we calculate a spin polarization and a resulting spin-torque, which are induced by {\it supercurrents} instead of resistive currents.  

Another aim of this paper is to study the topological structure in the Bogoliubov quasiparticle band in the heterostructure.  
We show that the band has a pair of Weyl points, and their positions are determined by the direction of magnetization. 
Consequently, inhomogeneous spin textures realize an effective magnetic field acting on Bogoliubov quasiparticles.   
Furthermore, the tilt of the Weyl cones can also be tuned by the strength of the SO coupling. 
Recently, {\it type-II} Weyl points, whose dispersion is strongly tilted so that it has a finite density of states at each Weyl point, are proposed  in materials such as LaAlGe~\cite{Xu2016}, MoTe$_2$~\cite{Huang2016,Xu2016a,Deng2016,Liang2016}, and WTe$_2$~\cite{Ali2014,Soluyanov2015,Wu2016}. They are predicted to have peculiar properties~\cite{Xu2015a, Yu2016, Volovik2016,Udagawa2016,Zyuzin2016, Koshino2016, Brien2016}, a chiral anomaly depending on the relative direction of the field and tilt\cite{Udagawa2016}, a nonanalytic behavior in the anomalous Hall effect~\cite{Zyuzin2016}, etc. 
For our superconducting system, we propose a realization of type-II Weyl points in the quasiparticle spectrum, and we show a transition between conventional (type-I) Weyl points and type-II Weyl points. Considering the effective magnetic field due to the spin texture, we expect intriguing transport properties to be realized in our system. 

The organization of this paper is as follows. In Sec.~II, we introduce a model, which describes a cubic chiral magnet with proximity-induced superconductivity. 
In Sec.~III, using a linear response theory, we calculate spin polarization induced by a supercurrent, which acts as a local spin torque that drives motions of skyrmions.   
In Sec.~IV, we study the band structure of Bogoliubov quasiparticles, and demonstrate how type-I (conventional) and type-II Weyl points appear. We also discuss the effect of inhomogeneity in spin textures. 
Section V is devoted to the summary and discussion. 

\begin{figure}
\includegraphics[width=6cm]{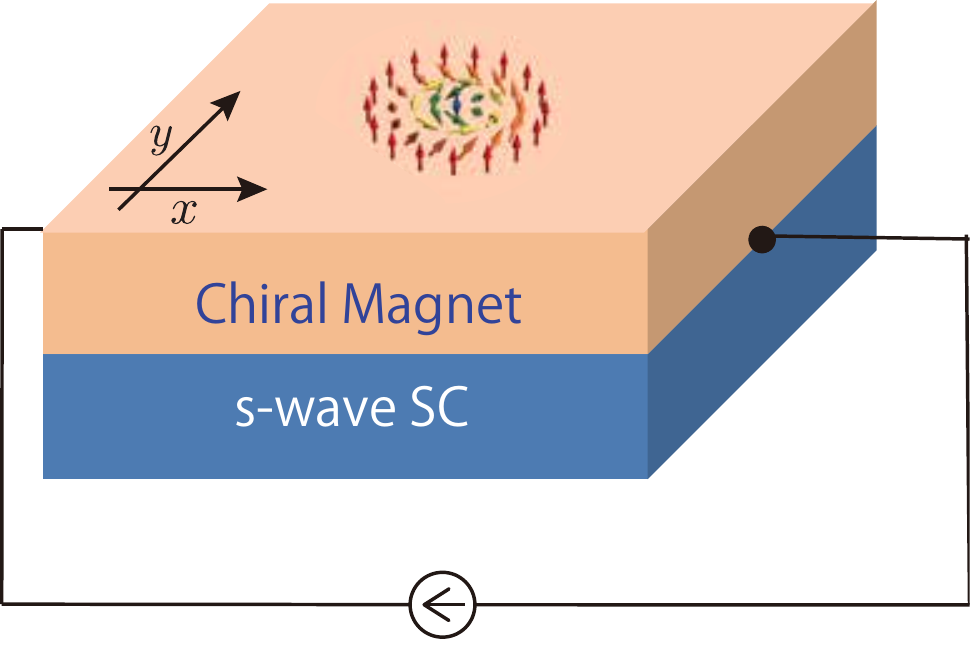}
\caption{
Heterostructure composed of a chiral magnet and an $s$-wave superconductor.  
}
\label{hetero}
\end{figure} 

\section{model}\label{model} 
We consider a heterostructure composed of an $s$-wave superconductor and a cubic chiral magnet (Fig.~\ref{hetero}), which is modeled by the following Hamiltonian:
\[
\mathcal H
&= \frac{1}{2}
\sum_{\bm k} 
\begin{pmatrix}
c^\dag_{\bm k }, 
c_{-\bm k}^{\rm T}
\end{pmatrix}
 H(\bm k)
\begin{pmatrix}
c_{\bm k } \\
(c_{-\bm k}^{\dag})^{\rm T}
\end{pmatrix},\]
where $c_{\bm k}^{\rm T}=(c_{\bm k \uparrow },c_{\bm k \downarrow}) $ are annihilation operators for electrons of spin up and down. 
The Bogoliubov-de Gennes (BdG) Hamiltonian $ H(\bm k)$ is
\[
 H(\bm k)=
\begin{pmatrix}
 H_0(\bm k) & -i\Delta \sigma_y \\
i\Delta^* \sigma_y &  -H_0(-\bm k) ^{\rm T}
\end{pmatrix},\label{original}
\]
with the normal part of the Hamiltonian 
\[
 H_0(\bm k)&= \left(\varepsilon_{\bm k} -\mu \right)\bm 1-J M \bm n \cdot \bm \sigma+\alpha_{so}\bm g(\bm k)\cdot \bm \sigma \label{normal},
\]
where $\bm \sigma$ are the Pauli matrices for spin space and  $\varepsilon_{\bm k}= -t \sum_{\ell=\{x, y,z\}} \cos ( k_\ell a )$ is a hopping energy with the lattice constant $a$ and band-width $2t$. The chemical potential is denoted by $\mu$, the exchange coupling coefficient is $J$, and the localized magnetic moment is denoted by $M \bm n$, with the unit vector $\bm n=(\cos\phi \sin\theta, \sin \phi \sin \theta,\cos \theta)$. We assume the  characteristic length of spatial variation to be much longer than the coherence length of superconductivity and the Fermi wave-length, so that $\bm{n}(\bm{r})$ can be regarded as being locally homogeneous. The effect of the inhomogeneity will be discussed in Sec.~\ref{quasi}.

The SO coupling in a cubic lattice model can be written as 
\[
 \bm g(\bm k)& = \left ( \sin (k_x a),\sin (k_y a),\sin (k_z a) \right),
\] 
which reflects the symmetry of cubic chiral magnets whose point-group symmetry is $T$. 
The proximity induced superconductivity is characterized by $\Delta$, where we take $s$-wave singlet pairings. 
The SO coupling supports the proximity of the superconductivity because it causes the tilt of the spin orientation in momentum space.   

\section{Supercurrent-induced spin polarization}
In normal metals or semiconductors, current induced magnetization dynamics has been studied extensively.  For nocollinear magnetic systems, one can use a spin transfer torque~\cite{Slonczewski1996, Slonczewski1999,Berger1996, Berger1999}, which originates from the transfer of spin angular momentum via spin-polarized currents. In noncentrosymmetric systems classified into gyrotropic materials, SO coupling gives rise to an additional current-induced torque, which is called the SO torque~\cite{Chernyshov2008,Manchon2008,Garate2009,Miron2010,Liu2012}. 
One way to interpret SO torques is as follows: in gyrotropic materials, charge currents induce a spin polarization\cite{Levitov1985, Edelstein1990,Fujimoto2012,Edelstein1989}, which is known as the Edelstein effect or the inverse spin-galvanic effect.  This induced spin polarization acts on the magnetization as a spin torque.  

On the other hand, much less is known in superconducting states. 
{Triplet pairing induced by spin flipping at the magnetic junction can realize superconducting counterpart of the spin transfer torques}~\cite{Braude2008, Linder2011}. 
Related to the SO torque, it has been discussed that supercurrents produce spin polarization with Rashba spin-orbit coupling in { paramagnetic} states~\cite{Edelstein1995,Yip2002,Yip2005,Fujimoto2005,Edelstein2005,Fujimoto2007}. This paper studies chiral magnets, where the exchange coupling is large. 
We show that the induced polarization depends on the direction of the magnetization.    

Using the BdG Hamiltonian [Eq.~\eqref{original}], one can calculate the spin polarization density of the system as   
\[
 s_\mu = \frac {1}{2} \frac{1}{V}\sum_{\bm k,n}n_F(\varepsilon_{n \bm k}) \langle n \bm k|\hat{S}_\mu|n \bm k\rangle,
\]
where $\ |n \bm k\rangle $ and $\varepsilon_{n \bm k}$ are the eigenstates and eigenenergy of the BdG Hamiltonian with the band indices $n =\{0,1,2,3\}$. $n_F(\varepsilon_{n \bm k})$ is the Fermi distribution function, and {$V$ is the volume}.  The spin operator is given by
\[
\hat S_\mu
&=\frac{\hbar}{2}\begin{pmatrix}
\sigma_{ \mu} & 0 \\
0&-\sigma_{\mu}^T \\
\end{pmatrix}. \label{spin}
\]

Now we consider a state with a finite supercurrent density $\bm j = -\frac{e}{m} n_s \hbar \bm Q$,
 where $m$ is an electron mass, $n_s$ is the superfluid density, $\bm Q$ is the shift of the Fermi surface, and $-e$ is the electron charge. With a perturbation theory, the leading correction to the polarization is obtained as
\[
\delta s_\mu =K_{\mu \nu}j_\nu,
\]
where 
\[
K_{\mu \nu}=& -\frac{m}{2e\hbar n_s} \lim_{\bm q \rightarrow 0} \frac{1}{V} \sum_{\bm k} \sum_{ n, m} \nonumber\\
&\times\frac{n_F(\varepsilon_{n \bm k}) - n_F(\varepsilon_{m \bm k+\bm q })  }{ \varepsilon_{m \bm k+\bm q }-\varepsilon_{n \bm k}  }
\langle n \bm k| \hat S_\mu| m \bm k +\bm q\rangle \langle m \bm k +\bm q| \hat V_\nu|n \bm k\rangle  \label{kubo}
\]
and the velocity operator $\hat V_\mu$ is defined by   
\[
\hat V_\mu = 
\begin{pmatrix}
\left.\frac{\partial  H_0}{\partial k_\mu}\right|_{\bm k}  & 0 \\
0&-\left( \left.\frac{\partial  H_0 }{\partial k_\mu} \right|_{-\bm k}\right)^{\rm T} 
\end{pmatrix}. 
\]

Equation~\eqref{kubo} is obtained by calculating the spin polarization using states with the center-of-mass momentum $\bm Q$ and expanding it in powers of  $\bm Q$ \cite{Maki1963,Edelstein1995}. (The derivation is given in Appendix \ref{deri}.)
In the rest of this section, we first study the response function $K_{\mu \nu}$ in two limiting cases of the parameters $JM$ and $\alpha_{so}$ and show the relative orientation between the induced magnetization and applied supercurrent. 
Second, we numerically calculate $K_{\mu \nu}$ with the original Hamiltonian [Eq.~\eqref{original}].  
Finally, we discuss the spin dynamics using the obtained spin torque and show that supercurrents can move a skyrmion.

\subsection{Limiting cases }\label{meeff}
In chiral magnets, due to the large exchange splitting,  $\alpha_{so} \ll JM$ is expected, but we first consider two limiting cases to provide an intuitive picture. In the limit of $JM/\alpha_{so} \rightarrow 0 $, the anisotropy due to the spin polarization $\bm n$ can be neglected. In this case, a symmetry argument straightforwardly leads to the relative orientation between $\delta \bm s$ and $\bm j$. 
The point-group symmetry $T$ leads to $K_{\mu \nu}=K \delta_{\mu \nu}$ \footnote{$C_2$ rotational symmetry in point group $T$ prohibits the off-diagonal components in $K_{\mu \nu}$, and $C_3$ rotational symmetry makes the diagonal components the same.}, and we obtain 
\[
\delta \bm s = K \bm j. \label{lim1}
\]
In this case, the spin polarization is induced in parallel to the supercurrent. This relation is in contrast to the case of Rashba SO coupling (e.g., $C_{4v}$) that gives $\delta \bm s \propto \bm {\hat z} \times \bm j$. 
 
In the limit of $ \alpha_{so} \ll JM$, $K_{\mu \nu}$ depends on the direction of $\bm n$. In particular, assuming $\{ JM, |\mu|\} \gg \{t, \alpha_{so},\Delta\}$, we perform a perturbative approach. 
 We first apply a unitary transformation $U(\bm k)$ on the BdG Hamiltonian [Eq.~\eqref{original}], which acts on both spin space and particle-hole space.  
 (The details are given in Appendix~\ref{unitary}.)  It diagonalizes the Hamiltonian up to the required order of a small parameter $\epsilon$ defined by $ \{t, \alpha_{so},\Delta \} \sim \epsilon JM  $, and we expand the Hamiltonian to the second order of $\epsilon$ as  
\[
 U^{\dag}(\bm k) H(\bm k) U(\bm k)= H'_0(\bm k)+ H'_1(\bm k), \]
where the diagonal matrix $ H'_0(\bm k)$ reads
\[
& H'_0(\bm k)=\text{diag}(E_{0 \bm k},E_{1\bm k},E_{2\bm k},E_{3\bm k}) \label{H0}
\]
with 
\[
E_{0\bm k}&= \varepsilon_{\bm k}-\mu-[JM- \alpha_{so}\bm g(\bm k)\cdot \bm n], \\
E_{1\bm k}&= \varepsilon_{\bm k}-\mu +[JM- \alpha_{so}\bm g(\bm k)\cdot \bm n], \\
E_{2\bm k}&=-( \varepsilon_{\bm k}-\mu)+[JM+\alpha_{so}\bm g(\bm k)\cdot \bm n], \\
E_{3\bm k}&=-(\varepsilon_{\bm k} -\mu) -[JM+\alpha_{so}\bm g(\bm k)\cdot \bm n],
\]
and every matrix element in $ H'_1(\bm k)$ is of the order of $\epsilon^2$.
For now we consider $\mu \sim -JM $ and project the Hilbert space to a space spanned by two eigenstates of $ H_0'(\bm k)$ with eigenenergies $ E_{0 \bm k} $ and $ E_{2\bm k} $, noting that $E_{3\bm k} \ll \{ E_{0\bm k} \sim E_{2 \bm k} \sim 0  \} \ll E_{1 \bm k}$. After the projection, we obtain an effective BdG Hamiltonian $  H'_0(\bm k)+ H'_1(\bm k) \mapsto H_{\rm eff}(\bm k) $, which is a $2\times 2$ matrix. 

We then  use the effective Hamiltonian $ H_{\rm eff} (\bm k)$ to calculate $K_{\mu \nu}$ [Eq.~\eqref{kubo}]. The detailed structure of the quasiparticle states is discussed in Sec.~\ref{quasi}. 
We define the eigenstates of $ H_{\rm eff} (\bm k)$ as $|| \pm,  \bm k \rangle$, whose eigenenergies are $\varepsilon_{\pm, \bm k}$. 
The spin and velocity operators in the projected space are defined as $ U^{\dag}(\bm k)  \hat S_\mu U(\bm k) \mapsto  \hat S_\mu^{\rm eff}$ and   $ U^{\dag}(\bm k)  \hat V_\mu U(\bm k) \mapsto \hat V_\mu^{\rm eff}$. (The details are given in Appendix~\ref{response}.)

For the calculation of a response against a supercurrent, the interband terms ($n\neq m$ in Eq.~\eqref{kubo}) are important.    
The directional dependence of $K_{\mu \nu}$ is obtained from the matrix element 
\[
 \mathcal S_{\mu \nu}& \equiv \langle -,\bm k||\hat S_\mu^{\rm eff} ||+, \bm k \rangle  \langle +, \bm k  ||\hat  V_\nu^{\rm eff} ||-,\bm k \rangle \label{matrix}\\
&=\frac{ a^2 \hbar(JM+|\mu|)^2 }{8(JM)^2 \mu^2}  \frac{ \alpha_{so}^3 |\Delta|^2 k_{\perp}^2 }{|\varepsilon_{+, \bm k}-\varepsilon_{-,\bm k}|^2 } n_{\mu}n_{\nu}\\
&=\langle +,\bm k  ||S_\mu^{\rm eff} ||-,\bm k \rangle \langle -,\bm k || V_\nu^{\rm eff} ||+,\bm k \rangle 
\]
to the lowest order of $\epsilon$, where we consider the momentum region that satisfies $|\bm k| a \ll1 $ and define $k_{\perp}^2=k^2-(\bm k \cdot \bm n)^2$. 
Integration over the momentum space as 
\[
K_{\mu \nu} 
= -\frac{m}{e\hbar n_s}  \int\frac{ d^3\bm k}{(2\pi)^3} \frac{n_F(\varepsilon_{-,\bm k})-n_F(\varepsilon_{+, \bm k}) }{\varepsilon_{+, \bm k}-\varepsilon_{-,\bm k} } \mathcal S_{\mu \nu }
\]
 leads to the relation $K_{\mu \nu} =K'  n_{\mu}n_\nu$; that is, the induced spin polarization is given by 
\[
\delta \bm s = K' \bm n (\bm n \cdot \bm j), \label{lim2}
\]
up to the lowest order in $\epsilon$.
The spin polarization is induced parallel to $\bm n$, and  it is not induced when the applied supercurrent is perpendicular to $\bm n$. 
As we will see later, this term does not produce a spin torque since the spin torque is given by $\bm n \times \delta \bm s$. 
In the higher order of $\epsilon$, other terms such as the form of Eq.~(\ref{lim1}) are expected to appear and lead to the spin torque, which is important for the dynamics of skyrmions discussed below.

\subsection{Numerical results} \label{mecal}
Next, we investigate the dependence of $K_{\mu \nu}$ on parameters and the direction of $\bm n$ for intermediate values of $\alpha_{so}/JM $ which interpolate the two limiting cases considered in the previous section. 
For this purpose, we numerically calculate $K_{\mu \nu}$ with the BdG Hamiltonian in Eq.~\eqref{original} 
and compare the results to the analytical asympotic results obtained in the last subsection. The main results in this section are as follows: (i) $K_{\mu \nu }$ depends on the direction of $\bm n$, and its dependence is well fitted by $K_{\mu \nu} = K^{(0)}\delta_{\mu \nu} +K^{(1)} n_{\mu} n_\nu$, which interpolates the asymptotic behaviors in the two limiting cases [Eqs.~\eqref{lim1} and \eqref{lim2}].  (ii) The numerical results for $\alpha_{so} /JM \ll 1$ confirm the asymptotic form given in Eq.~\eqref{lim2}. (iii) $K_{\mu \mu}$ monotonically increase with increasing $\alpha_{so}$, and $K_{\mu\mu}$ with $\mu$ in the direction parallel to $\bm n$
dominates over that with $\mu$ in the direction perpendicular to $\bm n$ for small $\alpha_{SO}/JM$, in accordance with Eq.~\eqref{lim2}. 

\begin{figure}
\includegraphics[width=7.3cm]{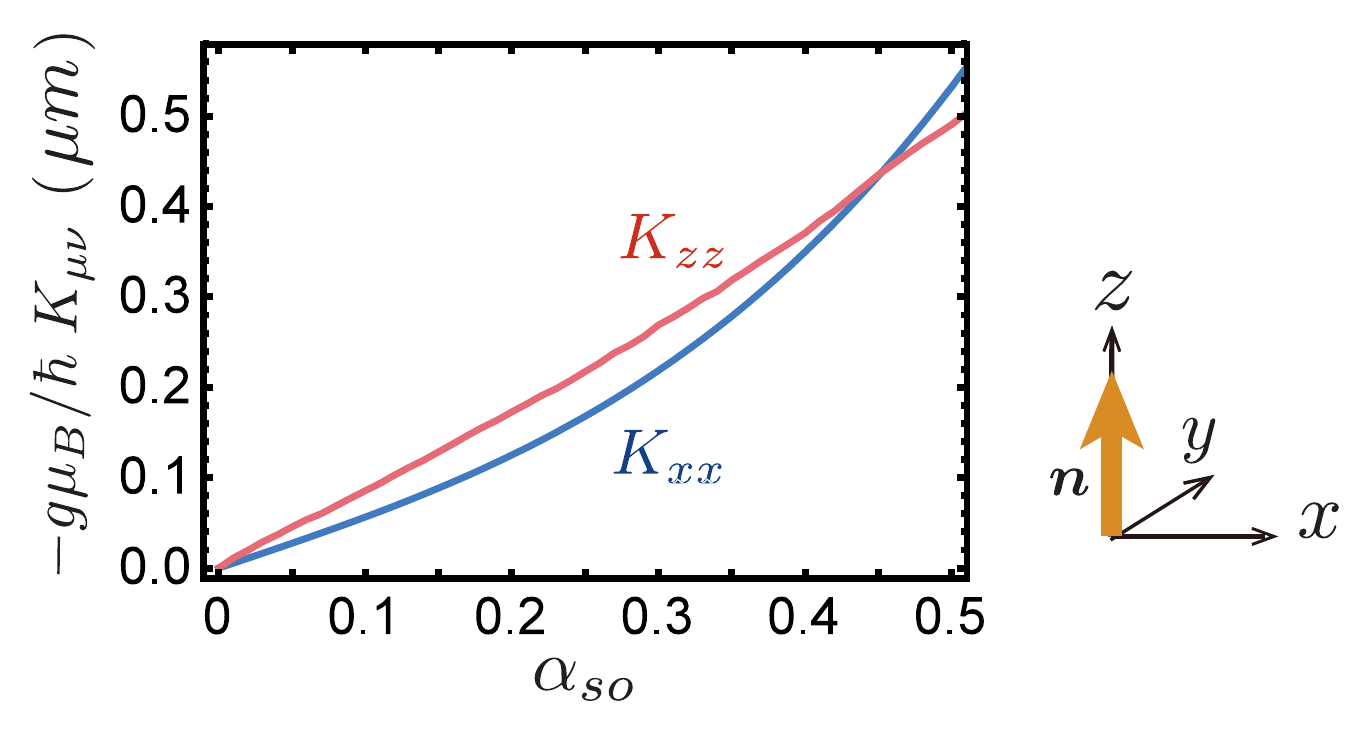}
\caption{
Dependence of $K_{xx}$ and $K_{zz}$ on the parameter $\alpha_{so}$ 
for $t=1.0,\ JM=2.0,\   \Delta =0.2,\  \mu =-4.8$, 
when the direction of the localized spin is $ \bm n= (0,0,1)$. 
Both $K_{xx}$ and $K_{zz}$ increase monotonically as $\alpha_{so}$ increases.  
In the limit of $ \alpha_{so}/JM \ll 1 $, the magnitude of  $K_{xx}$ is suppressed compared to $K_{zz}$, which is in agreement with the asymptotic form given by  Eq.~\eqref{lim2}.
}
\label{Kxx_Kzz}
\end{figure}

\begin{figure}
\includegraphics[width=7.cm]{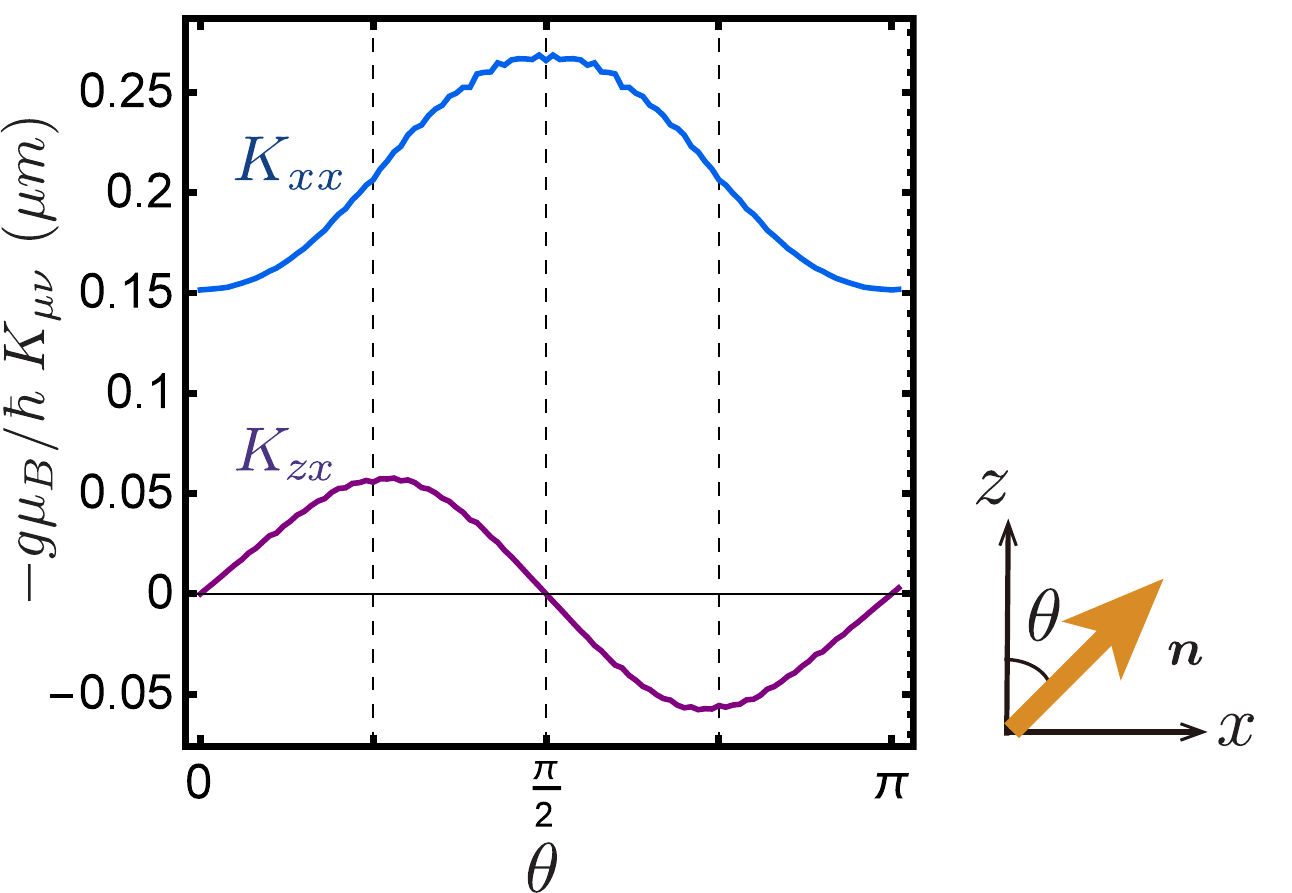}
\caption{
Dependence of $K_{xx}$ and $K_{zx}$ on the directional angle $\theta$ that parametrizes the localized spin as $\bm n= (\sin \theta,0,\cos\theta)$. 
The parameters are set as $t=1.0,\ JM=2.0,\   \Delta =0.2,\  \mu =-4.8$, and $\alpha_{so}=0.2$.
The diagonal component $K_{xx}$ has an offset value that is independent of $\theta$, and it is well fitted by $K_{xx}= 0.118 n_x^2 +0.150 $. The off-diagonal component, $K_{xy}$, is fitted by $K_{xy}=0.112 n_x n_y$. 
}
\label{Theta}
\end{figure}

Figures \ref{Kxx_Kzz} and \ref{Theta} show numerical results of the spin-polarization density per unit of supercurrent density with the parameters $t=1.0,\ JM=2.0,\   \Delta =0.2,\  \mu =-4.8$. 
The lattice constant is $a=10^{-1} nm$ and the London penetration depth $\lambda =50 nm$.     
In Fig.~\ref{Kxx_Kzz}, we show the dependence on $\alpha_{so}$ of $K_{xx}$ and $K_{zz}$ in the case with 
$\bm n=(0,0,1)$. 
$K_{xx}$ and $K_{zz}$ monotonically increase with increasing $\alpha_{so}$, and the difference between them arises from the directional anisotropy due to $\bm n$, which is consistent with the asymptotic from Eq.~\eqref{lim2} with $\bm n=(0,0,1)$; the magnitude of  $K_{xx}$ is suppressed compared to $K_{zz}$  in the limit of $\alpha_{so}/JM \ll 1 $. However, it is noted that, in contrast to Eq.~\eqref{lim2} which implies zero $K_{xx}$ for $\bm n=(0,0,1)$,
$K_{xx}$ is finite in Fig.~\ref{Kxx_Kzz} even for small values of $\alpha_{SO}$ because of the higher order terms neglected in Eq.~\eqref{lim2} and also the continuum approximation. 

In Fig.~\ref{Theta}, we 
show the dependence of $K_{xx}$ and $K_{zx}$ on the direction of $\bm n$ rotating in the $xz$ plane.
$\bm n$ is parametrized as $\bm n= (\sin \theta, 0, \cos\theta)$ with  $0 \leq \theta\leq \pi$. 
The magnitude of $K_{xx}$ and $K_{zx}$ varies depending on $\theta$, and their dependence is well fitted by $K_{\mu \nu} = K^{(0)}\delta_{\mu \nu} +K^{(1)} n_{\mu} n_\nu$, i.e. $K_{xx}=  K^{(0)}+K^{(1)} \sin^2 \theta$ and 
$K_{zx}= K^{(1)} \sin\theta \cos \theta$. 
Thus, the $\theta$-dependence of $K_{\mu \nu}$ is qualitatively given by the sum of  Eqs.~\eqref{lim1} and \eqref{lim2}.
In this calculation, we set  $\alpha_{so}=0.2$, and we note that  the  behaviors of the $\theta$-dependence are not qualitatively changed by varying the value of $\alpha_{so}$.
Other components such as $K_{xy}$ are calculated as well, and we conclude that, in general, the dependence of $K_{\mu \nu}$ on $\bm n$ is qualitatively
 given by the sum of  the two limiting cases, Eqs.~\eqref{lim1} and \eqref{lim2}, for intermediate values of $\alpha_{so}/JM $. 

\subsection {Equation of motion for collective coordinates}
Using the obtained results, we discuss skyrmion dynamics induced by supercurrents. 
{We consider a heterostucture system with a supercurrent flow.}  Here we neglect the effect of normal currents, which would be suppressed in the presence of superconductivity.  Our analysis is based on the Landau-Lifshitz-Gilbert(LLG) equation for the localized spin $\bm n$,
\begin{eqnarray}
\frac{d\bm{n}}{dt}=-\gamma \bm{n}\times \bm{H} +\alpha_G \bm{n}\times \frac{d\bm{n}}{dt}+\bm{T},
\label{eq:LLG}
\end{eqnarray}
where $\gamma$ is the gyromagnetic ratio, $\bm{H}$ is the effective magnetic field given by 
the derivative of the free energy with respect to the magnetization,
$\alpha_G $ is the Gilbert damping constant, 
and $\bm T$ is a supercurrent induced spin torque. 
We focus on a spin-orbit torque that originates from the spin polarization discussed in the last section. The spin polarization is given by the form $\delta \bm s = K^{(0)} \bm j + K^{(1)} \bm n (\bm n\cdot \bm j)$, and it results in the following spin torque:   
\[
\bm T &= \frac{JM}{ \hbar^2 } \bm n \times \delta \bm s ,\\
&= \frac{JM K^{(0)} }{ \hbar^2 } \bm n \times \bm j; \label{torque}
\]
that is., only $ K^{(0)} \bm j $ acts as a spin torque, and the anisotropy due to $\bm n$ does not affect the spin torque even for large $JM$ with the above form.  

The dynamics of skyrmions is well described by the equation of motion of collective coordinates of a texture \cite{Tretiakov2008, Everschor2012,Schulz2012a,Hals2014}. 
Neglecting the deformation of skyrmions, we take the center-of-mass coordinate $\bm R(t)$ as the collective coordinate. The spin texture is given by $\bm n(\bm r, t)=\bm n^{sk}(\bm r-\bm R(t) )$, where $\bm n^{sk}(\bm r)$ is the spin configuration with a skyrmion at the origin. $\bm n_{sk}(\bm r)$ has a finite topological number as  
\[
\frac{1}{4\pi}\int d^2 \bm r \bm n^{sk}\cdot \left(\frac{\partial \bm n^{sk}}{\partial x} \times\frac{\partial \bm n^{sk}}{\partial y} \right)=-1.
\] 
For simplicity, we consider a specific skyrmion configuration given by 
\[
n^{sk}_x(\bm r)&=-\sin \Theta(r)\frac{y }{r},\\
n^{sk}_y(\bm r)&=\sin \Theta(r) \frac{x }{r}.\\
n^{sk}_z(\bm r)&=\cos \Theta(r),
\]
where $\Theta(r)$ only depends on only $r=\sqrt{x^2+y^2}$, and it satisfies $\Theta(0)=\pi $ and $\Theta(\infty)=0$.  
With this configuration and the torque [Eq.~\eqref{torque}], we calculate the time dependence of $\bm R(t)$ from the LLG equation [Eq.~\eqref{eq:LLG}], and obtain 
\[
\dot{R}_x&=-\Lambda_0 j_x +\alpha_G \Lambda_1 j_y, \label{time1}\\
\dot{R}_y&=-\Lambda_0 j_y -\alpha_G \Lambda_1 j_x, \label{time2}
\]
where 
\[
\Lambda_0&= \frac{JM K^{(0)}}{\hbar^2}\frac{L_0}{4\pi},\\
\Lambda_1&= \frac{JM K^{(0)}}{\hbar^2}\frac{\Gamma_0L_0 }{16\pi^2}.
\]
$L_0$ and  $\Gamma_0$ depends on the function $\Theta(r)$; $L_0=  \int d^2\bm r \partial_x \bm n_y = -\int d^2\bm r \partial_y \bm n_x $ is of the order of the radius of a skyrmion, and $\Gamma_0=  \int d^2 \bm r \partial_x \bm n\cdot \partial_x \bm n =\int d^2 \bm r \partial_y \bm n\cdot \partial_y \bm n $.

The supercurrent-induced torque can give rise to the drift motion of skyrmions, and the finite Gilbert damping coefficient $\alpha_G$ gives the transverse motion against the supercurrent, which is a well-known effect for skyrmion dynamics~\cite{Iwasaki2013}.  Equations.~\eqref{time1} and \eqref{time2} are compatible with the  results obtained from phenomenological arguments in the presence of SO coupling~\cite{Hals2014}.  

\section{ Quasiparticle structure}\label{quasi}
In the heterostructure of interest, Bogoliubov quasiparticles have nontrivial band topology; a pair of Weyl points exists.  We first study the effective Hamiltonian which describes low-energy quasiparticles and then numerically demonstrate the existence of type-I and type-II Weyl points. At the end of this section, we discuss the effect of the spatial inhomogeneity in spin textures and show that an effective magnetic field acts on the quasiparticles.  

\subsection{Type-I and Type-II Weyl fermions}
The structure of Bogoliubov quasiparticles in the presence of spin textures such as skyrmions is rather complicated.
Before tackling this problem, we first consider the case of a homogeneous exchange field. 
This analysis is valid provided that the spatial variation of the spin texture is sufficiently weak compared to the Fermi wave-length and {the superconducting coherence length},
and thus the spin configuration is locally approximated by a homogeneous structure.

For a homogeneous spin configuration, the effective Hamitonian is derived in Sec.~\ref{meeff}.
In the limit of $\{ JM, |\mu|\} \gg \{t, \alpha_{so},\Delta\} $ with  $\mu \sim -JM$, the Hamiltonian for low-energy quasiparticles 
 is given by
\[
H_{\rm eff}(\bm k)=\alpha_{so} \bm k \cdot \bm n  \bm 1+ 
\begin{pmatrix}
\xi_{\rm  eff}(\bm k) & \Delta_{\rm  eff} \bm \gamma \cdot \bm k\\
\Delta_{\rm  eff} \bm \gamma^* \cdot \bm k & -\xi_{\rm  eff}(\bm k)\\
\end{pmatrix} \label{effH}
\]
where 
\[
\Delta_{ \rm eff}&=\frac{|\Delta| \alpha_{so} (JM +|\mu|) }{2JM |\mu|} ,\\
\xi_{\rm eff}(\bm k)&= \varepsilon_{\bm k} -JM-\mu -\frac{|\Delta|^2}{2\mu} +\frac{\alpha_{so}^2 }{2JM }k_\perp^2 .
\]
Here we consider the momentum $|\bm k| a \ll 1$ ($a=1$), and define $k_{\perp}^2=k^2-(\bm k \cdot \bm n)^2$. 
{
A complex vector  $\bm \gamma=(\gamma_x, \gamma_y, \gamma_z)$ satisfies $|{\rm Re} \bm \gamma|=|{\rm Im} \bm \gamma| = 1$ and ${\rm Re} \bm \gamma \times {\rm Im} \bm \gamma = -\bm n$. (See Appendix~\ref{unitary} for details)
}
The two bands cross at $\bm k=\pm k_0 \bm n$ if there exists a $k_0$ that satisfies $\xi_{\rm eff}(k_0 \bm n) =0$, and the crossing points are shifted from the zero energy by $\pm \alpha_{so} k_0$. 
These crossing points are protected by the spin rotation symmetry along $\bm n$;  that is, two crossing bands of interest have different eigenvalues of $\hat {\bm S}\cdot \bm n$, which commutes with $H ( k_0\bm n) $.

\begin{figure}[t]
\includegraphics[width=8.5cm]{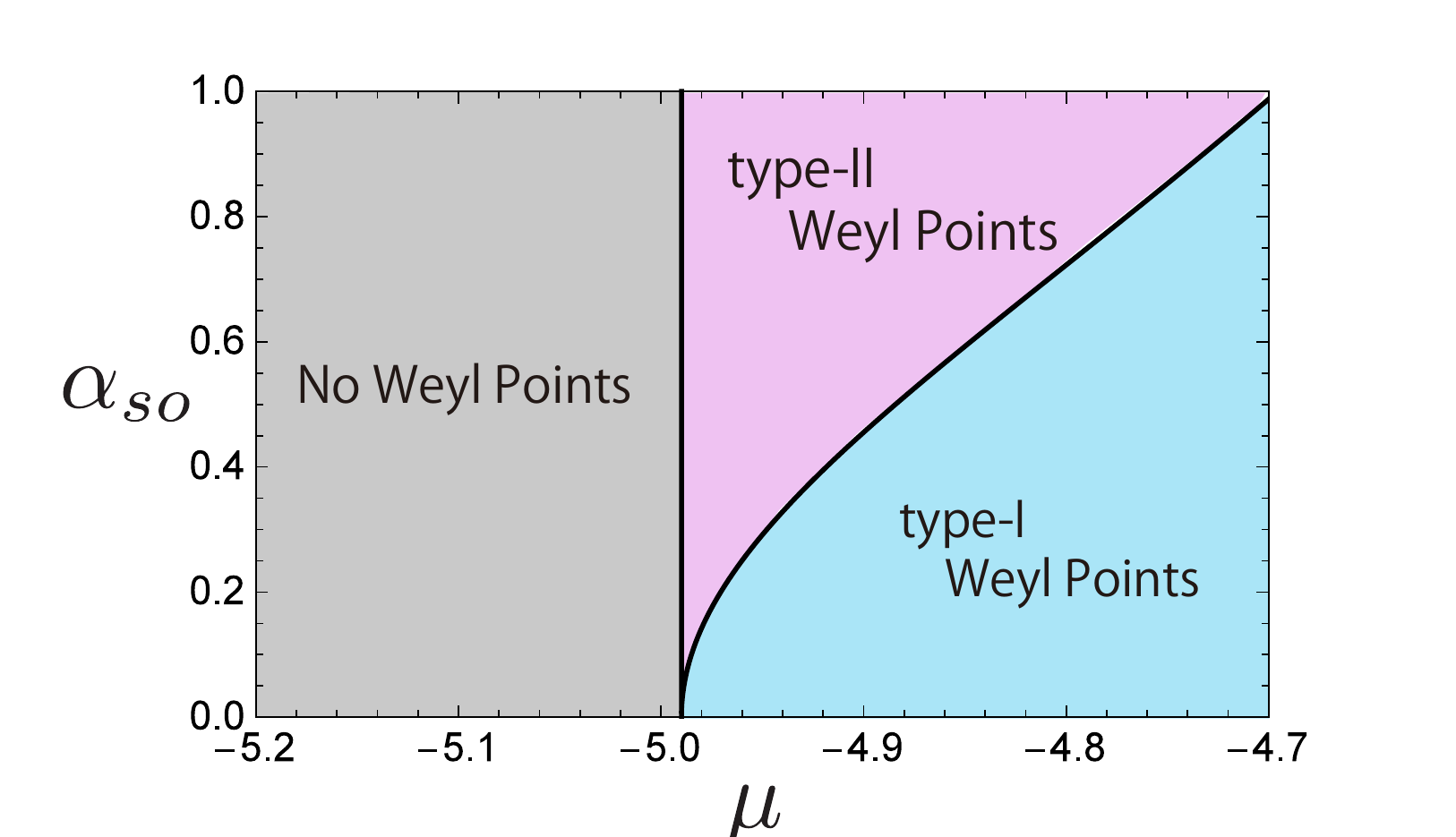}
\caption{Phase diagram of Weyl points as a function of chemical potential $\mu$ and the SO coupling coefficient $\alpha_{so}$ with the parameters $ t=1.0,\ JM=2.0,\ \Delta =0.2$.  
 In phases denoted by type-I (type-II) Weyl points, a pair of type-I (type-II) Weyl points exist.    
}
\label{Chern}
\end{figure}

\begin{figure}[tbh]
\includegraphics[width=7cm]{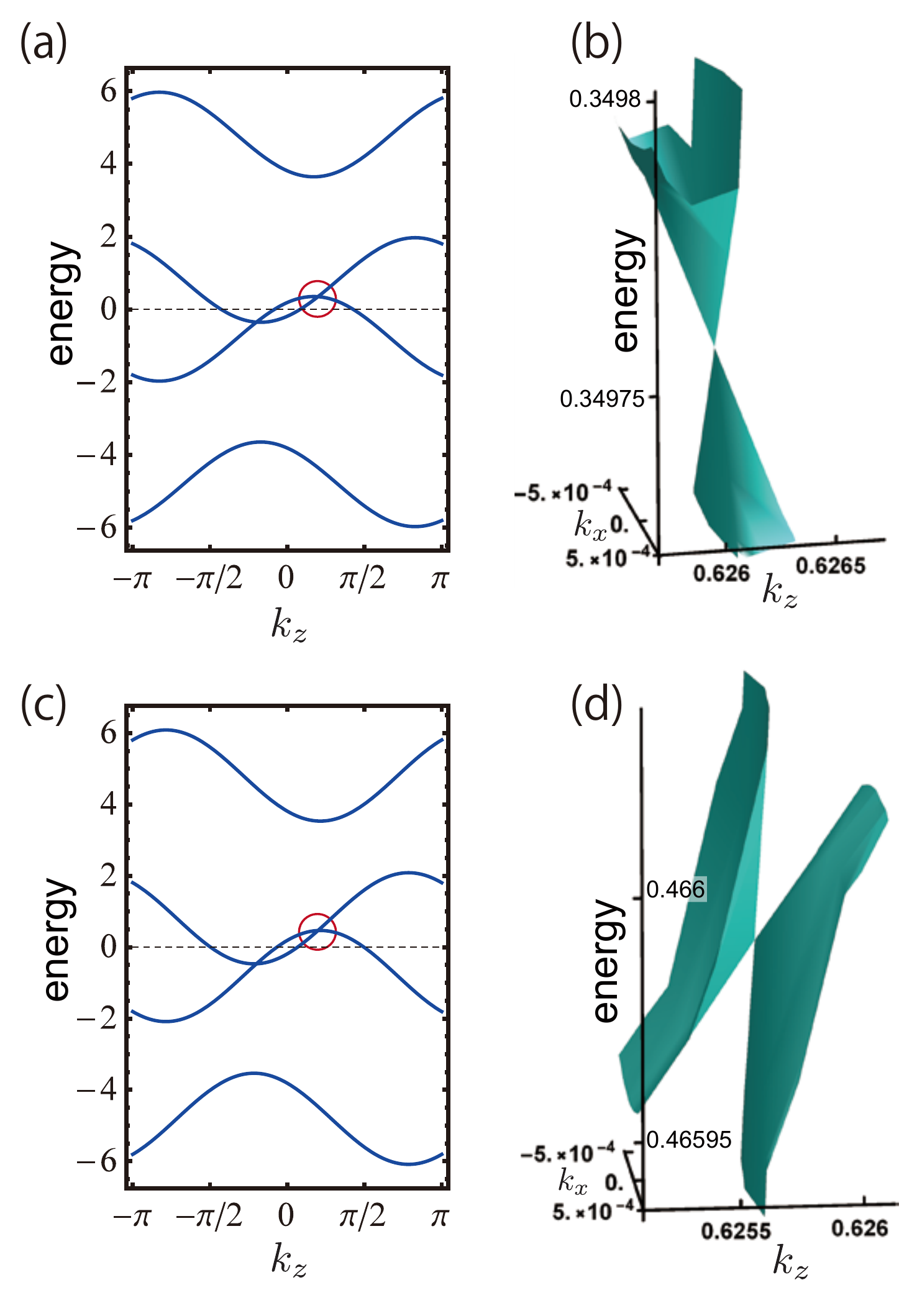}
\caption{Energy spectrum for the parameters $t=1.0,\ \Delta =0.2,\ JM=2.0,\ \mu=-4.8$ and $\bm n =(0,0,1)$. The Weyl points are at $k=(0,0,\pm k_0)$. 
(a) and (b) are the energy spectrum in the $k_x=k_y=0$ and in $k_z\ k_x$ plane with the parameter $\alpha_{so}=0.6 $.  A pair of type-I Weyl cones exists in the bands of quasiparticles  
(c) and (d) are the energy spectrum in the $k_x=k_y=0$ and in  $k_z\ k_x$ plane with the parameter $\alpha_{so}=0.8 $. 
 A pair of type-II Weyl cones exists. 
}
\label{typeII}
\end{figure}

Without loss of generality, we can define the $x, y $ and $z$ axes of momentum along ${\rm Re} \bm \gamma,{\rm Im} \bm \gamma,$ and $\bm n$ so that the crossing points locate at $k_0 \hat{\bm z}$. 
The effective Hamiltonian around the crossing points reads 
\[
&H_{\rm eff}(\pm k_0\hat{\bm z} +\bm q)=\nonumber\\
&\hspace{20pt}
\alpha_{so} (\pm k_0 +q_z ) \bm 1+ 
\begin{pmatrix}
\pm t k_0 q_z & \Delta_{\rm  eff} (q_x +i q_y)\\
\Delta_{\rm  eff} (q_x -i q_y) & \mp t k_0 q_z\\
\end{pmatrix}.  
\]
Each cone is tilted by $\alpha_{so}q_z$, and numerical calculations of the Berry curvature with the original BdG Hamiltonian [Eq.~\eqref{original}] show that they are a pair of Weyl points, which are a source and a sink of the Berry curvature.  
 The numerical result of the phase diagram is summarized in Fig.~\ref{Chern}.  
When the chemical potential is larger than the critical value, a pair of Weyl points with the opposite topological charge exists. 
In Figs.~\ref{typeII}(a) and (b), a band structure for the parameters $\mu =-4.8,$ and $ \alpha_{so}=0.6$ is shown, 
which has conventional (type-I)  Weyl cones. The dispersions are  weakly tilted in the $k_z$ direction.


Interestingly, changing $\alpha_{so}$ or $\mu$ leads to a transition between type-I and type-II Weyl cones. 
Type-II Weyl cones are characterized by the finite density of states at the crossing energy due to the tilt of the cones. The energy spectrum is shown in Fig.~\ref{typeII} (c) and (d) for $\mu =-4.8$ and $\alpha_{so} =0.6$. 
This tilt may induce distinct properties in transport or the Landau levels under an effective magnetic field discussed in the next section. 

We have shown the surface Majorana arc exists for both type-I and type-II Weyl points by numerical calculations with an open boundary.  Since the positions of the Weyl points are at $\pm k_0 \bm n$, the surface in which the Majorana arc appear can be changed by the direction of the magnetization. 

In the $k_z =0,\  \pm \pi$ plane, our system can be mapped to a well-known model.  By rotating the momentum by $90^\circ$ along the $k_z$ axis, the Hamiltonian [Eq.~\eqref{original}] is equivalent to a model to realize a spinless $p+ip$ superconductor with the Rashba SO coupling\cite{Sato2009, Sau2010}. These two-dimensional planes can have nonzero Chern number depending on the parameters, and the difference of the Chern number ensures the existence of  Weyl points.

\subsection{Inhomogeneous spin texture}
So far, we have studied a locally homogeneous system assuming that the spin texture varies weakly enough in space.   
Now we include the effect of inhomogeneity to the lowest order, and show the emergence of an effective magnetic field. 
We consider quasiparticle states around some arbitary point $\bm r_0$ in real space, and define $ \bm n_0 = \bm n(\bm r_0)$ and $\bm \gamma_0=\bm \gamma(\bm r_0)$.  We then rewrite as $\bm n(\bm r)= \bm n_0 +\delta \bm n(\bm r)$ and $\gamma(\bm r) = \gamma_0 +\delta \gamma(\bm r)$.
Around one of the Weyl point given by $\bm k=  k_0 \bm n_0 $, we define the Bogoliubov quasiparticle operator as $\psi(\bm r)=e^{  i k_0  \bm n_0\cdot \bm r } \tilde \psi_{+} (\bm r)$.  Up to the lowest order of spatial variation, the Hamiltonian for slowly varying field is given by   
{
\[
&H^+_{\rm eff}(\bm r )=
\alpha_{so} ( k_0 + \bm n_0 \cdot \hat{\bm p} ) \bm 1+ \nonumber\\
&\hspace{40pt}
\begin{pmatrix}
 t k_0 \bm n_0 \cdot \hat{\bm p} & \Delta_{\rm eff} \bm \gamma_0 \cdot (\hat{\bm p}-k_0  \bm n(\bm r) )\\
\Delta_{\rm eff} \bm \gamma_0^* \cdot (\hat{\bm p}-k_0  \bm n(\bm r) )& - t k_0   \bm n_0 \cdot \hat{\bm p} \\
\end{pmatrix}, \\
&\hspace{27pt}=
\alpha_{so} ( k_0 + \bm n_0 \cdot (\hat{\bm p}-k_0 \delta \bm n(\bm r) ) \bm 1+ \nonumber\\
&\hspace{40pt}
\begin{pmatrix}
 t k_0 \bm n_0 \cdot (\hat{\bm p}-k_0 \delta \bm n(\bm r) ) & \Delta_{\rm eff} \bm \gamma_0 \cdot (\hat{\bm p}-k_0 \delta \bm n(\bm r) )\\
\Delta_{\rm eff} \bm \gamma_0^* \cdot (\hat{\bm p}-k_0  \delta\bm n(\bm r) )& - t k_0   \bm n_0 \cdot  (\hat{\bm p}-k_0 \delta \bm n(\bm r) ) \\
\end{pmatrix}, 
\]
where we have used the relation $\bm n(\bm r)\cdot \bm \gamma(\bm r)=0$ and $\bm n_0 \cdot \delta \bm n(\bm r) =0$. 
Minimal coupling $\hat{\bm p}-k_0  \delta\bm n(\bm r) $ exhibits that the effect of the spatial variation of spins can be described by the  effective magnetic field given by $\bm B_{\rm eff}(\bm r)=k_0 \bm \nabla \times \bm n(\bm r) $. }
Around the other Weyl point at $-k_0 \bm n_0$, the sign is the opposite: $\bm B_{\rm eff}(\bm r)=-k_0 \bm \nabla \times \bm n(\bm r) $. 

This result shows that Bogoliubov quasiparticles around each Weyl cone may form the Landau levels when $\bm \nabla \times \bm n(\bm r) \neq \bm 0$, and the chiral zero modes are expected to appear for each cone, which is studied in the A phase of $^3$He \cite{Combescot1986,Volovik1998, Bevan1997}.
The velocities of the chiral zero modes have the same direction for two cones to preserve the particle hole symmetry, and they are along 
the direction of $\bm \nabla \times \bm n(\bm r)$.

A skyrmion texture gives $\bm \nabla \times \bm n(\bm r) \neq \bm 0 $, and a skyrmion flow is expected to cause an effective electric field given by $\mp k_0 \partial_t \bm n$, which acts on quasiparticles in each cone. As has been discussed in  $^3$He \cite{Volovik1998, Bevan1997}, the quasiparticle excitation in the chiral zero mode is expected to cause a force acting on the skyrmion, which is perpendicular to the velocity of the skyrmion \footnote{This is distinguished from the Magnus force, which does not require the excitation of quasiparticles. }; that is., the anomaly of Weyl fermions can affect the skyrmion dynamics. 

Tuning the tilt may change the property of the chiral zero modes, as is studied in Weyl semimetals~\cite{Udagawa2016}. We leave the detailed study of skyrmion dynamics associated with Weyl fermions in chiral magnet for future work.  

\section{Summary and Discussion}
In this paper, we have studied a spin-torque induced by a supercurrent in a heterostructure composed of a cubic chiral magnet and an $s$-wave superconductor. With the numerical and analytical calculation, we have derived  the spin polarization induced by a supercurrent, which depends on the direction of the localized spin. The equation for the time evolution of skyrmion has been obtained.  
  
We have also pointed out the existence of a pair of Weyl points in the quasiparticle bands. The positions of the Weyl points are determined by the magnetization, and the resulting effective electromagnetic field may give rise to novel phenomena in spintronics. The tilt of the cones can also be changed by the strength of the SO coupling, and type-II Weyl points can be realized. We believe that the nontrivial band topology of Bogoliubov quasiparticles provides a new perspective in superconducting spintronics. 

We leave the discussion of the stability of the proximity induced superconducting gap for future work. The quantitative estimation requires self-consistent calculations of the superconducting gap. At this point, we expect that there should be a finite region where the superconducting gap is proximity-induced in the chiral magnet
when the interface between the superconductor and the chiral magnet is sufficiently smooth.

\appendix 
\section{Derivation of Eq.~\eqref{kubo} }\label{deri}
A state with a supercurrent can be modeled by the finite phase gradient of superconducting order parameter as 
\[
H_{SC}&= \frac{1}{2} \sum_{i}e^{2i \bm Q\cdot \bm r_i} c^{\dag}_i  (i\Delta \sigma_y) \left( c^{\dag}_i \right)^T +H.c. , 
\]
where $c_i =\frac{1}{\sqrt V }\sum_{\bm k} c_{\bm k}e^{i \bm k \cdot \bm r_i} $. 
To simplify the calculation, we introduce fermion operators $\tilde c_i = c_i e^{- i \bm Q\cdot \bm r_i}$. With this operator, the Hamiltonian can be rewritten as 
\[
\mathcal H
&= \frac{1}{2}
\sum_{\bm k} 
\begin{pmatrix}
\tilde c^\dag_{\bm k }, 
\tilde c_{-\bm k}^{\rm T}
\end{pmatrix}
 \tilde H(\bm k)
\begin{pmatrix}
\tilde c_{\bm k } \\
(\tilde c_{-\bm k}^{\dag})^{\rm T}
\end{pmatrix},\\
\tilde H(\bm k)&=
\begin{pmatrix}
 H_0(\bm k+\bm Q) &  -i\Delta \sigma_y \\
i\Delta^* \sigma_y &  -H_0(-\bm k+\bm Q) ^{\rm T}
\end{pmatrix},\\
&= H(\bm k)+\hat V_{\nu } Q_\nu +O(Q^2),
\]
where $\hat V_{\nu }$ is defined in Eq.~(9). 

Since the spin-polarization density is a local observable, it can be calculated with $\tilde c_{\bm k}$ as 
\[
s_\mu &= \frac{\hbar}{2V} \sum_{\bm k}\langle c^\dag_{\bm k} \sigma_\mu c_{\bm k}\rangle,\\
&= \frac{\hbar}{2V} \sum_{\bm k}\langle \tilde c^\dag_{\bm k} \sigma_\mu \tilde c_{\bm k}\rangle.
\]
A perturbation theory with $\tilde H(\bm k)$ leads to Eq.~(8).

\section{Effective Hamiltonian } \label{unitary}
In this appendix, we show the derivation of the effective Hamiltonian when $\{ JM, |\mu|\} \gg \{t, \alpha_{so},\Delta\} $.
We start from the BdG Hamiltonian in Eq.~\eqref{original} . 
Then we apply a unitary transformation $U^\dag(\bm k) H(\bm k)U(\bm k)$. The unitary matrix is given by 
\[
U(\bm k)=
\begin{pmatrix}
U_0&0\\
0&U_0^*
\end{pmatrix}
e^{iQ(\bm k)},
\]
where $U_0 $ rotates the spin axis as $U_0^\dag \bm n \cdot \bm \sigma U_0=\sigma_z$. 
\begin{widetext}
$Q(\bm k)$ is Hermite matrix given by 
\[
Q(\bm k)=\begin{pmatrix}
0&-\frac{i \alpha_{so}}{2M} \omega &0&- \frac{i\Delta}{2\mu}\\
\frac{i \alpha_{so}}{2M}  \omega^* &0& \frac{i\Delta}{2\mu} &0\\
0& -\frac{i\Delta ^*}{2\mu} &0&\frac{i \alpha_{so}}{2M} \omega^* \\
\frac{i\Delta ^*}{2\mu} &0&\frac{-i \alpha_{so}}{2M}\omega  &0
\end{pmatrix}
\]
with  
\[
 \omega = \sum_{\ell=x,y,z}\sin (k_\ell a) \left(U_0^\dag \sigma_\ell U_0\right)_{12}.
\]
Then we expand the Hamiltonian in the series of $\epsilon$ where we define the small parameter by $ \{t, \alpha_{so},\Delta \} \sim \epsilon JM $,  as
 $ U^{\dag}(\bm k) H(\bm k) U(\bm k)= H'_0(\bm k)+ H'_1(\bm k)+\mathcal O(\epsilon^3)$,
where 
$H'_0(\bm k)$ is given by Eq.~\eqref{H0}
and
\[
& H_1'(\bm k)=\begin{pmatrix}
 -\frac{|\Delta|^2}{2\mu}-\frac{\alpha_{so}^2 }{2JM}|\omega|^2&
 \frac{\alpha_{so}^2}{JM} \beta \omega&
u_0{ \Delta \omega  } & 
\frac{\Delta}{\mu}(\varepsilon_{\bm k} +\alpha_{so}\beta)\\
\frac{\alpha_{so}^2 }{JM}\beta\omega^* &
 -\frac{|\Delta|^2}{2\mu} + \frac{\alpha_{so}^2 }{2JM} |\omega|^2&
-\frac{\Delta}{\mu}(\varepsilon_{\bm k} -\alpha_{so}\beta )&
u_1\Delta \omega^* \\
u_0 \Delta^* \omega^*&
-\frac{\Delta^*}{\mu}(\varepsilon_{\bm k} -\alpha_{so}\beta )&
 \frac{|\Delta|^2}{2\mu}+ \frac{\alpha_{so}^2 }{2JM}|\omega|^2&
-\frac{\alpha_{so}^2}{JM}\beta\omega^* \\
\frac{\Delta^* }{\mu}(\varepsilon_{\bm k} +\alpha_{so}\beta)&
u_1\Delta^* \omega&
- \frac{\alpha_{so}^2}{JM}\beta\omega&
\frac{|\Delta|^2}{2\mu}-\frac{\alpha_{so}^2 }{2JM}| \omega|^2
\end{pmatrix},
\]
with $ \beta= \sum_{\ell=x,y,z}\sin (k_\ell a) n_\ell$, $u_0 =-\frac{\alpha_{so}(JM-\mu)} {2JM\mu } $, and $u_1=\frac{\alpha_{so} (JM+\mu) }{2JM\mu}$. When $\mu \sim -JM$, the eigenstates of $H'_0(\bm k)$ are energetically separated as $E_{3\bm k} \ll \{ E_{0\bm k} \sim E_{2 \bm k} \sim 0\} \ll E_{1 \bm k}$.  We project the Hilbert space to a space around zero energy, i.e. a space spanned by eigenstates with eigenenergy $E_{0 \bm k}$ and $E_{2 \bm k}$.  
\end{widetext}
{
The effective Hamiltonian $H_{\rm eff} (\bm k)$ is given by 
\[
H_{\rm eff} (\bm k)
&= \begin{pmatrix}
[H_0'(\bm k)]_{11} +[H_1'(\bm k)]_{11}  &[H_0'(\bm k)]_{13} +[H_1'(\bm k)]_{13} \\
[H_0'(\bm k)]_{31}+[H_1'(\bm k)]_{31} &[H_0'(\bm k)]_{33}+[H_1'(\bm k)]_{33} \\
\end{pmatrix},\\
&= 
\begin{pmatrix}
E_{0 \bm k}  -\frac{|\Delta|^2}{2\mu}-\frac{\alpha_{so}^2 }{2JM}|\omega|^2 & -\frac{\alpha_{so}|\Delta | (JM-\mu)} {2JM\mu } e^{i \varphi}\omega   \\
-\frac{\alpha_{so}|\Delta| (JM-\mu)}{2JM\mu } e^{-i \varphi} \omega^* & E_{2 \bm k } +\frac{|\Delta|^2}{2\mu}+ \frac{\alpha_{so}^2 }{2JM}|\omega|^2
\end{pmatrix},
\]
where we define $\Delta =|\Delta|e^{i \varphi}$ and $(H_n'(\bm k))_{ij}$ is the $(i,j)$ component of  $H_n'(\bm k)$. }
We introduce a complex vector $\bm \gamma=(\gamma_x, \gamma_y,\gamma_z)$ as
\[
\bm \gamma &= e^{i \varphi} \left(U_0^\dag \bm \sigma U_0\right)_{12}, 
\]
where $ \left(U_0^\dag \bm \sigma U_0\right)_{12}$ is the $(1,2)$ component in the $2 \times 2$ matrix $U_0^\dag \bm \sigma U_0 $.    
One can show the following relations  
\[ 
|{\rm Re} \bm \gamma|=| {\rm Im} \bm \gamma| &= 1,\\
 {\rm Re} \bm \gamma \cdot {\rm Im} \bm \gamma &=0,\\
{\rm Re} \bm \gamma \times{\rm Im} \bm \gamma &= -\bm n,\\
\epsilon_{\mu \nu \lambda} \partial_\mu (\bm \gamma^* \cdot \partial_\nu \bm \gamma)&= -2 i \epsilon_{\mu \nu \lambda}\bm n \cdot (\partial_\mu \bm n \times \partial_\nu \bm n).
\]
The last equation shows that the rotation of the superfluid velocity is given by the skyrmion density.  
Around the $\Gamma$ point $(|\bm k| a \ll 1)$, we obtain the effective Hamiltonian in Eq.~\eqref{effH}, noting that  
 $e^{i \varphi} \omega=\bm \gamma \cdot \bm k a$ and  $|\omega|^2=a^2k^2 -a^2(\bm k  \cdot \bm n)^2 $.  
 
\section{Edelstein effect with the effective Hamiltonian}\label{response}
In this section, we calculate the matrix element $\mathcal S_{\mu \nu}$ (Eq.~\eqref{matrix}) with the effective Hamiltonian in the continuum limit ($|\bm k| a \ll1$). 
The effective Hamiltonian (Eq.~\eqref{effH}) is
\[
H_{\rm eff}(\bm k) = f_0 \bm 1 + \bm f \cdot \bm \tau, 
\]
with $\bm \tau=(\tau_x, \tau_y, \tau_z)$ are pauli matrices, and
\[
f_0&= \alpha_{so}\bm k \cdot \bm n,\\
\bm f &=(f_1,f_2,f_3),\\
f_1&=\Delta_{\rm eff} {\rm Re}\bm{\gamma}\cdot \bm k,\\
f_2&=-\Delta_{\rm eff}{\rm Im}\bm{\gamma}\cdot \bm k,\\ 
f_3&=\xi_{\rm eff}(\bm k).
\]
We define $\bm f/|\bm f| = (\sin \theta_0 \cos \phi_0,\sin \theta_0 \sin \phi_0, \cos \theta_0 ) $, and then the eigenstates of Eq.~\eqref{effH} are given by 
\[
|| +, \bm k\rangle =\begin{pmatrix}
\cos \frac{\theta_0 }{2} e^{- i \phi_0} \\   
\sin \frac{\theta_0}{2}
\end{pmatrix},
\\
||-, \bm k \rangle =
\begin{pmatrix}
\sin \frac{\theta_0}{2}e^{- i \phi_0} \\   
-\cos \frac{\theta_0}{2}
\end{pmatrix}.
\]
The corresponding eigenstates are $\varepsilon_{\pm , \bm k} =f_0 \pm | \bm f |$, which depend on $k^2$ and $\bm k \cdot \bm n$. 
The spin and velocity operators in the projected space are 
\[
\hat S^{\rm eff}_\mu \sim
&\frac{\hbar}{2}
\left(
n_\mu \tau_z 
-\frac{\alpha_{so}}{M} 
{\rm Re} (\gamma_\mu^* \bm \gamma \cdot \bm k) 
\bm 1 \right),\\
\hat V^{\rm eff}_\mu \sim &
-t k_\mu \bm 1 +\alpha_{so}n_\mu \tau_z,
\] 
where we have expanded in the series of $\epsilon$. 
Thus we obtain 
\[
\mathcal S_{\mu \nu} 
& =\langle -,\bm k ||\hat S^{\rm eff}_\mu||+, \bm k \rangle \langle +, \bm k  ||\hat V^{\rm eff}_\nu ||-,\bm k \rangle ,\\
&=\langle -,\bm k ||\tau_z||+, \bm k \rangle \langle +, \bm k  || \tau_z ||-,\bm k \rangle \frac{\hbar}{2} \alpha_{so} n_{\mu}n_{\nu} ,\\
&=\sin^2 \theta_0  \frac{\hbar}{2} \alpha_{so} n_{\mu}n_{\nu},\\
&=\frac{\Delta_{\rm eff}^2 k_\perp^2}{\xi_{\rm eff}(\bm k)^2+\Delta_{\rm eff}^2k_\perp^2}\frac{\hbar}{2} \alpha_{so} n_{\mu}n_{\nu},
\]
where $k_{\perp}^2= k^2 -(\bm k \cdot \bm n)^2$.  
One can also show that $\langle +,\bm k ||S^{\rm eff}_\mu||-, \bm k \rangle \langle -, \bm k  ||V^{\rm eff}_\nu ||+,\bm k \rangle =\mathcal S_{\mu \nu}$.
\begin{acknowledgments}
We thank Muhammad Shahbaz Anwar, Alexander Balatsky, Akito Daido, Takuya Nomoto, Masatoshi Sato, Yuki Shiomi, and Yoichi Yanase for fruitful discussions. 
 This work was supported by the Grant-in-Aid for Scientific
Research from MEXT of Japan [Grants No. 23540406, No. 25220711, and No. 15H05852 (KAKENHI on Innovative Areas  “Topological Materials Science")]. 
R.T. is supported by a Japan Society for the Promotion of Science Fellowship for Young Scientists.
\end{acknowledgments}

\bibliographystyle{apsrev4-1}

\begin{thebibliography}{87}%
\makeatletter
\providecommand \@ifxundefined [1]{%
 \@ifx{#1\undefined}
}%
\providecommand \@ifnum [1]{%
 \ifnum #1\expandafter \@firstoftwo
 \else \expandafter \@secondoftwo
 \fi
}%
\providecommand \@ifx [1]{%
 \ifx #1\expandafter \@firstoftwo
 \else \expandafter \@secondoftwo
 \fi
}%
\providecommand \natexlab [1]{#1}%
\providecommand \enquote  [1]{``#1''}%
\providecommand \bibnamefont  [1]{#1}%
\providecommand \bibfnamefont [1]{#1}%
\providecommand \citenamefont [1]{#1}%
\providecommand \href@noop [0]{\@secondoftwo}%
\providecommand \href [0]{\begingroup \@sanitize@url \@href}%
\providecommand \@href[1]{\@@startlink{#1}\@@href}%
\providecommand \@@href[1]{\endgroup#1\@@endlink}%
\providecommand \@sanitize@url [0]{\catcode `\\12\catcode `\$12\catcode
  `\&12\catcode `\#12\catcode `\^12\catcode `\_12\catcode `\%12\relax}%
\providecommand \@@startlink[1]{}%
\providecommand \@@endlink[0]{}%
\providecommand \url  [0]{\begingroup\@sanitize@url \@url }%
\providecommand \@url [1]{\endgroup\@href {#1}{\urlprefix }}%
\providecommand \urlprefix  [0]{URL }%
\providecommand \Eprint [0]{\href }%
\providecommand \doibase [0]{http://dx.doi.org/}%
\providecommand \selectlanguage [0]{\@gobble}%
\providecommand \bibinfo  [0]{\@secondoftwo}%
\providecommand \bibfield  [0]{\@secondoftwo}%
\providecommand \translation [1]{[#1]}%
\providecommand \BibitemOpen [0]{}%
\providecommand \bibitemStop [0]{}%
\providecommand \bibitemNoStop [0]{.\EOS\space}%
\providecommand \EOS [0]{\spacefactor3000\relax}%
\providecommand \BibitemShut  [1]{\csname bibitem#1\endcsname}%
\let\auto@bib@innerbib\@empty
\bibitem [{\citenamefont {Hasan}\ and\ \citenamefont {Kane}(2010)}]{Hasan2010}%
  \BibitemOpen
  \bibfield  {author} {\bibinfo {author} {\bibfnamefont {M.~Z.}\ \bibnamefont
  {Hasan}}\ and\ \bibinfo {author} {\bibfnamefont {C.~L.}\ \bibnamefont
  {Kane}},\ }\href {\doibase 10.1103/RevModPhys.82.3045} {\bibfield  {journal}
  {\bibinfo  {journal} {Rev. Mod. Phys.}\ }\textbf {\bibinfo {volume} {82}},\
  \bibinfo {pages} {3045} (\bibinfo {year} {2010})}\BibitemShut {NoStop}%
\bibitem [{\citenamefont {Qi}\ and\ \citenamefont {Zhang}(2011)}]{Qi2011}%
  \BibitemOpen
  \bibfield  {author} {\bibinfo {author} {\bibfnamefont {X.~L.}\ \bibnamefont
  {Qi}}\ and\ \bibinfo {author} {\bibfnamefont {S.~C.}\ \bibnamefont {Zhang}},\
  }\href {\doibase 10.1103/RevModPhys.83.1057} {\bibfield  {journal} {\bibinfo
  {journal} {Rev. Mod. Phys.}\ }\textbf {\bibinfo {volume} {83}},\ \bibinfo
  {pages} {1057} (\bibinfo {year} {2011})}\BibitemShut {NoStop}%
\bibitem [{\citenamefont {Volovik}(2003)}]{Volovik2003}%
  \BibitemOpen
  \bibfield  {author} {\bibinfo {author} {\bibfnamefont {G.~E.}\ \bibnamefont
  {Volovik}},\ }\href@noop {} {\emph {\bibinfo {title} {{The Universe in a
  Helium Droplet}}}}\ (\bibinfo  {publisher} {Oxford University Press},\
  \bibinfo {address} {Oxford},\ \bibinfo {year} {2003})\BibitemShut {NoStop}%
\bibitem [{\citenamefont {Murakami}(2007)}]{Murakami2007}%
  \BibitemOpen
  \bibfield  {author} {\bibinfo {author} {\bibfnamefont {S.}~\bibnamefont
  {Murakami}},\ }\href {\doibase 10.1088/1367-2630/9/9/356} {\bibfield
  {journal} {\bibinfo  {journal} {New J. Phys.}\ }\textbf {\bibinfo {volume}
  {9}},\ \bibinfo {pages} {356} (\bibinfo {year} {2007})}\BibitemShut {NoStop}%
\bibitem [{\citenamefont {Wan}\ \emph {et~al.}(2011)\citenamefont {Wan},
  \citenamefont {Turner}, \citenamefont {Vishwanath},\ and\ \citenamefont
  {Savrasov}}]{Wan2011}%
  \BibitemOpen
  \bibfield  {author} {\bibinfo {author} {\bibfnamefont {X.}~\bibnamefont
  {Wan}}, \bibinfo {author} {\bibfnamefont {A.~M.}\ \bibnamefont {Turner}},
  \bibinfo {author} {\bibfnamefont {A.}~\bibnamefont {Vishwanath}}, \ and\
  \bibinfo {author} {\bibfnamefont {S.~Y.}\ \bibnamefont {Savrasov}},\ }\href
  {\doibase 10.1103/PhysRevB.83.205101} {\bibfield  {journal} {\bibinfo
  {journal} {Phys. Rev. B}\ }\textbf {\bibinfo {volume} {83}},\ \bibinfo
  {pages} {205101} (\bibinfo {year} {2011})}\BibitemShut {NoStop}%
\bibitem [{\citenamefont {Xu}\ \emph {et~al.}(2011)\citenamefont {Xu},
  \citenamefont {Weng}, \citenamefont {Wang}, \citenamefont {Dai},\ and\
  \citenamefont {Fang}}]{Xu2011}%
  \BibitemOpen
  \bibfield  {author} {\bibinfo {author} {\bibfnamefont {G.}~\bibnamefont
  {Xu}}, \bibinfo {author} {\bibfnamefont {H.}~\bibnamefont {Weng}}, \bibinfo
  {author} {\bibfnamefont {Z.}~\bibnamefont {Wang}}, \bibinfo {author}
  {\bibfnamefont {X.}~\bibnamefont {Dai}}, \ and\ \bibinfo {author}
  {\bibfnamefont {Z.}~\bibnamefont {Fang}},\ }\href {\doibase
  10.1103/PhysRevLett.107.186806} {\bibfield  {journal} {\bibinfo  {journal}
  {Phys. Rev. Lett.}\ }\textbf {\bibinfo {volume} {107}},\ \bibinfo {pages}
  {186806} (\bibinfo {year} {2011})}\BibitemShut {NoStop}%
\bibitem [{\citenamefont {Yang}\ \emph {et~al.}(2011)\citenamefont {Yang},
  \citenamefont {Lu},\ and\ \citenamefont {Ran}}]{Yang2011b}%
  \BibitemOpen
  \bibfield  {author} {\bibinfo {author} {\bibfnamefont {K.~-Y.}\ \bibnamefont
  {Yang}}, \bibinfo {author} {\bibfnamefont {Y.~-M.}\ \bibnamefont {Lu}}, \ and\
  \bibinfo {author} {\bibfnamefont {Y.}~\bibnamefont {Ran}},\ }\href {\doibase
  10.1103/PhysRevB.84.075129} {\bibfield  {journal} {\bibinfo  {journal} {Phys.
  Rev. B}\ }\textbf {\bibinfo {volume} {84}},\ \bibinfo {pages} {075129} (\bibinfo {year} {2011})},\BibitemShut {NoStop}%
\bibitem [{\citenamefont {Burkov}\ and\ \citenamefont
  {Balents}(2011)}]{Burkov2011}%
  \BibitemOpen
  \bibfield  {author} {\bibinfo {author} {\bibfnamefont {A.~A.}\ \bibnamefont
  {Burkov}}\ and\ \bibinfo {author} {\bibfnamefont {L.}~\bibnamefont
  {Balents}},\ }\href {\doibase 10.1103/PhysRevLett.107.127205} {\bibfield
  {journal} {\bibinfo  {journal} {Phys. Rev. Lett.}\ }\textbf {\bibinfo
  {volume} {107}},\ \bibinfo {pages} {127205} (\bibinfo {year}
  {2011})}\BibitemShut {NoStop}%
\bibitem [{\citenamefont {Hal{\'{a}}sz}\ and\ \citenamefont
  {Balents}(2012)}]{Halasz2012}%
  \BibitemOpen
  \bibfield  {author} {\bibinfo {author} {\bibfnamefont {G.~B.}\ \bibnamefont
  {Hal{\'{a}}sz}}\ and\ \bibinfo {author} {\bibfnamefont {L.}~\bibnamefont
  {Balents}},\ }\href {\doibase 10.1103/PhysRevB.85.035103} {\bibfield
  {journal} {\bibinfo  {journal} {Phys. Rev. B}\ }\textbf {\bibinfo {volume}
  {85}},\ \bibinfo {pages} {035103} (\bibinfo {year} {2012})}\BibitemShut
  {NoStop}%
\bibitem [{\citenamefont {Witczak-Krempa}\ and\ \citenamefont
  {Kim}(2012)}]{Witczak-Krempa2012}%
  \BibitemOpen
  \bibfield  {author} {\bibinfo {author} {\bibfnamefont {W.}~\bibnamefont
  {Witczak-Krempa}}\ and\ \bibinfo {author} {\bibfnamefont {Y.~B.}\
  \bibnamefont {Kim}},\ }\href {\doibase 10.1103/PhysRevB.85.045124} {\bibfield
   {journal} {\bibinfo  {journal} {Phys. Rev. B}\ }\textbf {\bibinfo {volume}
  {85}},\ \bibinfo {pages} {045124} (\bibinfo {year} {2012})}\BibitemShut
  {NoStop}%
\bibitem [{\citenamefont {Hosur}\ \emph {et~al.}(2012)\citenamefont {Hosur},
  \citenamefont {Parameswaran},\ and\ \citenamefont {Vishwanath}}]{Hosur2012}%
  \BibitemOpen
  \bibfield  {author} {\bibinfo {author} {\bibfnamefont {P.}~\bibnamefont
  {Hosur}}, \bibinfo {author} {\bibfnamefont {S.~A.}\ \bibnamefont
  {Parameswaran}}, \ and\ \bibinfo {author} {\bibfnamefont {A.}~\bibnamefont
  {Vishwanath}},\ }\href {\doibase 10.1103/PhysRevLett.108.046602} {\bibfield
  {journal} {\bibinfo  {journal} {Phys. Rev. Lett.}\ }\textbf {\bibinfo
  {volume} {108}},\ \bibinfo {pages} {046602} (\bibinfo {year}
  {2012})}\BibitemShut {NoStop}%
\bibitem [{\citenamefont {Zyuzin}\ and\ \citenamefont
  {Burkov}(2012)}]{Zyuzin2012}%
  \BibitemOpen
  \bibfield  {author} {\bibinfo {author} {\bibfnamefont {A.~A.}\ \bibnamefont
  {Zyuzin}}\ and\ \bibinfo {author} {\bibfnamefont {A.~A.}\ \bibnamefont
  {Burkov}},\ }\href {\doibase 10.1103/PhysRevB.86.115133} {\bibfield
  {journal} {\bibinfo  {journal} {Phys. Rev. B}\ }\textbf {\bibinfo {volume}
  {86}},\ \bibinfo {pages} {115133} (\bibinfo {year} {2012})}\BibitemShut
  {NoStop}%
\bibitem [{\citenamefont {Aji}(2012)}]{Aji2012}%
  \BibitemOpen
  \bibfield  {author} {\bibinfo {author} {\bibfnamefont {V.}~\bibnamefont
  {Aji}},\ }\href {\doibase 10.1103/PhysRevB.85.241101} {\bibfield  {journal}
  {\bibinfo  {journal} {Phys. Rev. B}\ }\textbf {\bibinfo {volume} {85}},\
  \bibinfo {pages} {241101} (\bibinfo {year} {2012})}\BibitemShut {NoStop}%
\bibitem [{\citenamefont {Son}\ and\ \citenamefont {Yamamoto}(2012)}]{Son2012}%
  \BibitemOpen
  \bibfield  {author} {\bibinfo {author} {\bibfnamefont {D.~T.}\ \bibnamefont
  {Son}}\ and\ \bibinfo {author} {\bibfnamefont {N.}~\bibnamefont {Yamamoto}},\
  }\href {\doibase 10.1103/PhysRevLett.109.181602} {\bibfield  {journal}
  {\bibinfo  {journal} {Phys. Rev. Lett.}\ }\textbf {\bibinfo {volume} {109}},\
  \bibinfo {pages} {181602} (\bibinfo {year} {2012})}\BibitemShut {NoStop}%
\bibitem [{\citenamefont {Liu}\ \emph {et~al.}(2013)\citenamefont {Liu},
  \citenamefont {Ye},\ and\ \citenamefont {Qi}}]{Liu2013}%
  \BibitemOpen
  \bibfield  {author} {\bibinfo {author} {\bibfnamefont {C.-X.}\ \bibnamefont
  {Liu}}, \bibinfo {author} {\bibfnamefont {P.}~\bibnamefont {Ye}}, \ and\
  \bibinfo {author} {\bibfnamefont {X.-L.}\ \bibnamefont {Qi}},\ }\href
  {\doibase 10.1103/PhysRevB.87.235306} {\bibfield  {journal} {\bibinfo
  {journal} {Phys. Rev. B}\ }\textbf {\bibinfo {volume} {87}},\ \bibinfo
  {pages} {235306} (\bibinfo {year} {2013})}\BibitemShut {NoStop}%
\bibitem [{\citenamefont {R{\"{o}}ssler}\ \emph {et~al.}(2006)\citenamefont
  {R{\"{o}}ssler}, \citenamefont {Bogdanov},\ and\ \citenamefont
  {Pfleiderer}}]{Rossler2006}%
  \BibitemOpen
  \bibfield  {author} {\bibinfo {author} {\bibfnamefont {U.~K.}\ \bibnamefont
  {R{\"{o}}ssler}}, \bibinfo {author} {\bibfnamefont {A.~N.}\ \bibnamefont
  {Bogdanov}}, \ and\ \bibinfo {author} {\bibfnamefont {C.}~\bibnamefont
  {Pfleiderer}},\ }\href {\doibase 10.1038/nature05056} {\bibfield  {journal}
  {\bibinfo  {journal} {Nature (London)}\ }\textbf {\bibinfo {volume} {442}},\ \bibinfo
  {pages} {797} (\bibinfo {year} {2006})}\BibitemShut {NoStop}%
\bibitem [{\citenamefont {M{\"{u}}hlbauer}\ \emph {et~al.}(2009)\citenamefont
  {M{\"{u}}hlbauer}, \citenamefont {Binz}, \citenamefont {Jonietz},
  \citenamefont {Pfleiderer}, \citenamefont {Rosch}, \citenamefont {Neubauer},
  \citenamefont {Georgii},\ and\ \citenamefont {B{\"{o}}ni}}]{Muhlbauer2009}%
  \BibitemOpen
  \bibfield  {author} {\bibinfo {author} {\bibfnamefont {S.}~\bibnamefont
  {M{\"{u}}hlbauer}}, \bibinfo {author} {\bibfnamefont {B.}~\bibnamefont
  {Binz}}, \bibinfo {author} {\bibfnamefont {F.}~\bibnamefont {Jonietz}},
  \bibinfo {author} {\bibfnamefont {C.}~\bibnamefont {Pfleiderer}}, \bibinfo
  {author} {\bibfnamefont {A.}~\bibnamefont {Rosch}}, \bibinfo {author}
  {\bibfnamefont {A.}~\bibnamefont {Neubauer}}, \bibinfo {author}
  {\bibfnamefont {R.}~\bibnamefont {Georgii}}, \ and\ \bibinfo {author}
  {\bibfnamefont {P.}~\bibnamefont {B{\"{o}}ni}},\ }\href {\doibase
  10.1126/science.1166767} {\bibfield  {journal} {\bibinfo  {journal}
  {Science}\ }\textbf {\bibinfo {volume} {323}},\ \bibinfo {pages} {915}
  (\bibinfo {year} {2009})}\BibitemShut {NoStop}%
\bibitem [{\citenamefont {Yu}\ \emph {et~al.}(2010)\citenamefont {Yu},
  \citenamefont {Onose}, \citenamefont {Kanazawa}, \citenamefont {Park},
  \citenamefont {Han}, \citenamefont {Matsui}, \citenamefont {Nagaosa},\ and\
  \citenamefont {Tokura}}]{Yu2010a}%
  \BibitemOpen
  \bibfield  {author} {\bibinfo {author} {\bibfnamefont {X.~Z.}\ \bibnamefont
  {Yu}}, \bibinfo {author} {\bibfnamefont {Y.}~\bibnamefont {Onose}}, \bibinfo
  {author} {\bibfnamefont {N.}~\bibnamefont {Kanazawa}}, \bibinfo {author}
  {\bibfnamefont {J.~H.}\ \bibnamefont {Park}}, \bibinfo {author}
  {\bibfnamefont {J.~H.}\ \bibnamefont {Han}}, \bibinfo {author} {\bibfnamefont
  {Y.}~\bibnamefont {Matsui}}, \bibinfo {author} {\bibfnamefont
  {N.}~\bibnamefont {Nagaosa}}, \ and\ \bibinfo {author} {\bibfnamefont
  {Y.}~\bibnamefont {Tokura}},\ }\href {\doibase 10.1038/nature09124}
  {\bibfield  {journal} {\bibinfo  {journal} {Nature (London)}\ }\textbf {\bibinfo
  {volume} {465}},\ \bibinfo {pages} {901} (\bibinfo {year}
  {2010})}\BibitemShut {NoStop}%
\bibitem [{\citenamefont {M{\"{u}}nzer}\ \emph {et~al.}(2010)\citenamefont
  {M{\"{u}}nzer}, \citenamefont {Neubauer}, \citenamefont {Adams},
  \citenamefont {M{\"{u}}hlbauer}, \citenamefont {Franz}, \citenamefont
  {Jonietz}, \citenamefont {Georgii}, \citenamefont {B{\"{o}}ni}, \citenamefont
  {Pedersen}, \citenamefont {Schmidt}, \citenamefont {Rosch},\ and\
  \citenamefont {Pfleiderer}}]{Munzer2010}%
  \BibitemOpen
  \bibfield  {author} {\bibinfo {author} {\bibfnamefont {W.}~\bibnamefont
  {M{\"{u}}nzer}}, \bibinfo {author} {\bibfnamefont {A.}~\bibnamefont
  {Neubauer}}, \bibinfo {author} {\bibfnamefont {T.}~\bibnamefont {Adams}},
  \bibinfo {author} {\bibfnamefont {S.}~\bibnamefont {M{\"{u}}hlbauer}},
  \bibinfo {author} {\bibfnamefont {C.}~\bibnamefont {Franz}}, \bibinfo
  {author} {\bibfnamefont {F.}~\bibnamefont {Jonietz}}, \bibinfo {author}
  {\bibfnamefont {R.}~\bibnamefont {Georgii}}, \bibinfo {author} {\bibfnamefont
  {P.}~\bibnamefont {B{\"{o}}ni}}, \bibinfo {author} {\bibfnamefont
  {B.}~\bibnamefont {Pedersen}}, \bibinfo {author} {\bibfnamefont
  {M.}~\bibnamefont {Schmidt}}, \bibinfo {author} {\bibfnamefont
  {A.}~\bibnamefont {Rosch}}, \ and\ \bibinfo {author} {\bibfnamefont
  {C.}~\bibnamefont {Pfleiderer}},\ }\href {\doibase
  10.1103/PhysRevB.81.041203} {\bibfield  {journal} {\bibinfo  {journal} {Phys.
  Rev. B}\ }\textbf {\bibinfo {volume} {81}},\ \bibinfo {pages} {041203}
  (\bibinfo {year} {2010})}\BibitemShut {NoStop}%
\bibitem [{\citenamefont {Yu}\ \emph {et~al.}(2011)\citenamefont {Yu},
  \citenamefont {Kanazawa}, \citenamefont {Onose}, \citenamefont {Kimoto},
  \citenamefont {Zhang}, \citenamefont {Ishiwata}, \citenamefont {Matsui},\
  and\ \citenamefont {Tokura}}]{Yu2011}%
  \BibitemOpen
  \bibfield  {author} {\bibinfo {author} {\bibfnamefont {X.~Z.}\ \bibnamefont
  {Yu}}, \bibinfo {author} {\bibfnamefont {N.}~\bibnamefont {Kanazawa}},
  \bibinfo {author} {\bibfnamefont {Y.}~\bibnamefont {Onose}}, \bibinfo
  {author} {\bibfnamefont {K.}~\bibnamefont {Kimoto}}, \bibinfo {author}
  {\bibfnamefont {W.~Z.}\ \bibnamefont {Zhang}}, \bibinfo {author}
  {\bibfnamefont {S.}~\bibnamefont {Ishiwata}}, \bibinfo {author}
  {\bibfnamefont {Y.}~\bibnamefont {Matsui}}, \ and\ \bibinfo {author}
  {\bibfnamefont {Y.}~\bibnamefont {Tokura}},\ }\href {\doibase
  10.1038/nmat2916} {\bibfield  {journal} {\bibinfo  {journal} {Nat. Mater.}\
  }\textbf {\bibinfo {volume} {10}},\ \bibinfo {pages} {106} (\bibinfo {year}
  {2011})}\BibitemShut {NoStop}%
\bibitem [{\citenamefont {Nagaosa}\ and\ \citenamefont
  {Tokura}(2013)}]{Nagaosa2013}%
  \BibitemOpen
  \bibfield  {author} {\bibinfo {author} {\bibfnamefont {N.}~\bibnamefont
  {Nagaosa}}\ and\ \bibinfo {author} {\bibfnamefont {Y.}~\bibnamefont
  {Tokura}},\ }\href {\doibase 10.1038/nnano.2013.243} {\bibfield  {journal}
  {\bibinfo  {journal} {Nat. Nanotechnol.}\ }\textbf {\bibinfo {volume} {8}},\
  \bibinfo {pages} {899} (\bibinfo {year} {2013})}\BibitemShut {NoStop}%
\bibitem [{\citenamefont {Stone}(1996)}]{Stone1996}%
  \BibitemOpen
  \bibfield  {author} {\bibinfo {author} {\bibfnamefont {M.}~\bibnamefont
  {Stone}},\ }\href {\doibase 10.1103/PhysRevB.53.16573} {\bibfield  {journal}
  {\bibinfo  {journal} {Phys. Rev. B}\ }\textbf {\bibinfo {volume} {53}},\
  \bibinfo {pages} {16573} (\bibinfo {year} {1996})}\BibitemShut {NoStop}%
\bibitem [{\citenamefont {Volovik}(1987)}]{Volovik1987}%
  \BibitemOpen
  \bibfield  {author} {\bibinfo {author} {\bibfnamefont {G.}~\bibnamefont
  {Volovik}},\ }\href {http://iopscience.iop.org/0022-3719/20/7/003} {\bibfield
   {journal} {\bibinfo  {journal} {J. Phys. C Solid State Phys.}\ }\textbf
  {\bibinfo {volume} {20}},\ \bibinfo {pages} {L83} (\bibinfo {year}
  {1987})}\BibitemShut {NoStop}%
\bibitem [{\citenamefont {Jonietz}\ \emph {et~al.}(2010)\citenamefont
  {Jonietz}, \citenamefont {M{\"{u}}hlbauer}, \citenamefont {Pfleiderer},
  \citenamefont {Neubauer}, \citenamefont {M{\"{u}}nzer}, \citenamefont
  {Bauer}, \citenamefont {Adams}, \citenamefont {Georgii}, \citenamefont
  {B{\"{o}}ni}, \citenamefont {Duine}, \citenamefont {Everschor}, \citenamefont
  {Garst},\ and\ \citenamefont {Rosch}}]{Jonietz2010}%
  \BibitemOpen
  \bibfield  {author} {\bibinfo {author} {\bibfnamefont {F.}~\bibnamefont
  {Jonietz}}, \bibinfo {author} {\bibfnamefont {S.}~\bibnamefont
  {M{\"{u}}hlbauer}}, \bibinfo {author} {\bibfnamefont {C.}~\bibnamefont
  {Pfleiderer}}, \bibinfo {author} {\bibfnamefont {A.}~\bibnamefont
  {Neubauer}}, \bibinfo {author} {\bibfnamefont {W.}~\bibnamefont
  {M{\"{u}}nzer}}, \bibinfo {author} {\bibfnamefont {A.}~\bibnamefont {Bauer}},
  \bibinfo {author} {\bibfnamefont {T.}~\bibnamefont {Adams}}, \bibinfo
  {author} {\bibfnamefont {R.}~\bibnamefont {Georgii}}, \bibinfo {author}
  {\bibfnamefont {P.}~\bibnamefont {B{\"{o}}ni}}, \bibinfo {author}
  {\bibfnamefont {R.~A.}\ \bibnamefont {Duine}}, \bibinfo {author}
  {\bibfnamefont {K.}~\bibnamefont {Everschor}}, \bibinfo {author}
  {\bibfnamefont {M.}~\bibnamefont {Garst}}, \ and\ \bibinfo {author}
  {\bibfnamefont {A.}~\bibnamefont {Rosch}},\ }\href {\doibase
  10.1126/science.1195709} {\bibfield  {journal} {\bibinfo  {journal}
  {Science}\ }\textbf {\bibinfo {volume} {330}},\ \bibinfo {pages} {1648}
  (\bibinfo {year} {2010})}\BibitemShut {NoStop}%
\bibitem [{\citenamefont {Yu}\ \emph {et~al.}(2012)\citenamefont {Yu},
  \citenamefont {Kanazawa}, \citenamefont {Zhang}, \citenamefont {Nagai},
  \citenamefont {Hara}, \citenamefont {Kimoto}, \citenamefont {Matsui},
  \citenamefont {Onose},\ and\ \citenamefont {Tokura}}]{Yu2012a}%
  \BibitemOpen
  \bibfield  {author} {\bibinfo {author} {\bibfnamefont {X.}~\bibnamefont
  {Yu}}, \bibinfo {author} {\bibfnamefont {N.}~\bibnamefont {Kanazawa}},
  \bibinfo {author} {\bibfnamefont {W.}~\bibnamefont {Zhang}}, \bibinfo
  {author} {\bibfnamefont {T.}~\bibnamefont {Nagai}}, \bibinfo {author}
  {\bibfnamefont {T.}~\bibnamefont {Hara}}, \bibinfo {author} {\bibfnamefont
  {K.}~\bibnamefont {Kimoto}}, \bibinfo {author} {\bibfnamefont
  {Y.}~\bibnamefont {Matsui}}, \bibinfo {author} {\bibfnamefont
  {Y.}~\bibnamefont {Onose}}, \ and\ \bibinfo {author} {\bibfnamefont
  {Y.}~\bibnamefont {Tokura}},\ }\href {\doibase 10.1038/ncomms1990} {\bibfield
   {journal} {\bibinfo  {journal} {Nat. Commun.}\ }\textbf {\bibinfo {volume}
  {3}},\ \bibinfo {pages} {988} (\bibinfo {year} {2012})}\BibitemShut {NoStop}%
\bibitem [{\citenamefont {Iwasaki}\ \emph {et~al.}(2013)\citenamefont
  {Iwasaki}, \citenamefont {Mochizuki},\ and\ \citenamefont
  {Nagaosa}}]{Iwasaki2013}%
  \BibitemOpen
  \bibfield  {author} {\bibinfo {author} {\bibfnamefont {J.}~\bibnamefont
  {Iwasaki}}, \bibinfo {author} {\bibfnamefont {M.}~\bibnamefont {Mochizuki}},
  \ and\ \bibinfo {author} {\bibfnamefont {N.}~\bibnamefont {Nagaosa}},\ }\href
  {\doibase 10.1038/nnano.2013.176} {\bibfield  {journal} {\bibinfo  {journal}
  {Nat. Nanotechnol.}\ }\textbf {\bibinfo {volume} {8}},\ \bibinfo {pages}
  {742} (\bibinfo {year} {2013})}\BibitemShut {NoStop}%
\bibitem [{\citenamefont {Fert}\ \emph {et~al.}(2013)\citenamefont {Fert},
  \citenamefont {Cros},\ and\ \citenamefont {Sampaio}}]{Fert2013}%
  \BibitemOpen
  \bibfield  {author} {\bibinfo {author} {\bibfnamefont {A.}~\bibnamefont
  {Fert}}, \bibinfo {author} {\bibfnamefont {V.}~\bibnamefont {Cros}}, \ and\
  \bibinfo {author} {\bibfnamefont {J.}~\bibnamefont {Sampaio}},\ }\href
  {\doibase 10.1038/nnano.2013.29} {\bibfield  {journal} {\bibinfo  {journal}
  {Nat. Nanotechnol.}\ }\textbf {\bibinfo {volume} {8}},\ \bibinfo {pages}
  {152} (\bibinfo {year} {2013})}\BibitemShut {NoStop}%
\bibitem [{\citenamefont {Lin}\ \emph {et~al.}(2013)\citenamefont {Lin},
  \citenamefont {Reichhardt}, \citenamefont {Batista},\ and\ \citenamefont
  {Saxena}}]{Lin2013}%
  \BibitemOpen
  \bibfield  {author} {\bibinfo {author} {\bibfnamefont {S.-Z.}\ \bibnamefont
  {Lin}}, \bibinfo {author} {\bibfnamefont {C.}~\bibnamefont {Reichhardt}},
  \bibinfo {author} {\bibfnamefont {C.~D.}\ \bibnamefont {Batista}}, \ and\
  \bibinfo {author} {\bibfnamefont {A.}~\bibnamefont {Saxena}},\ }\href
  {\doibase 10.1103/PhysRevB.87.214419} {\bibfield  {journal} {\bibinfo
  {journal} {Phys. Rev. B}\ }\textbf {\bibinfo {volume} {87}},\ \bibinfo
  {pages} {214419} (\bibinfo {year} {2013})}\BibitemShut {NoStop}
\bibitem [{\citenamefont {Schutte}\ \emph {et~al.}(2014)\citenamefont
  {Schutte}, \citenamefont {Iwasaki}, \citenamefont {Rosch},\ and\
  \citenamefont {Nagaosa}}]{Schutte2014c}%
  \BibitemOpen
  \bibfield  {author} {\bibinfo {author} {\bibfnamefont {C.}~\bibnamefont
  {Schutte}}, \bibinfo {author} {\bibfnamefont {J.}~\bibnamefont {Iwasaki}},
  \bibinfo {author} {\bibfnamefont {A.}~\bibnamefont {Rosch}}, \ and\ \bibinfo
  {author} {\bibfnamefont {N.}~\bibnamefont {Nagaosa}},\ }\href {\doibase
  10.1103/PhysRevB.90.174434} {\bibfield  {journal} {\bibinfo  {journal} {Phys.
  Rev. B}\ }\textbf {\bibinfo {volume} {90}},\ \bibinfo {pages} {174434}
  (\bibinfo {year} {2014})}\BibitemShut {NoStop}%
\bibitem [{\citenamefont {Slonczewski}(1996)}]{Slonczewski1996}%
  \BibitemOpen
  \bibfield  {author} {\bibinfo {author} {\bibfnamefont {J.}~\bibnamefont
  {Slonczewski}},\ }\href {\doibase 10.1016/0304-8853(96)00062-5} {\bibfield
  {journal} {\bibinfo  {journal} {J. Magn. Magn. Mater.}\ }\textbf {\bibinfo
  {volume} {159}},\ \bibinfo {pages} {L1} (\bibinfo {year} {1996})}\BibitemShut
  {NoStop}%
\bibitem [{\citenamefont {Slonczewski}(1999)}]{Slonczewski1999}%
  \BibitemOpen
  \bibfield  {author} {\bibinfo {author} {\bibfnamefont {J.}~\bibnamefont
  {Slonczewski}},\ }\href {\doibase 10.1016/S0304-8853(99)00043-8} {\bibfield
  {journal} {\bibinfo  {journal} {J. Magn. Magn. Mater.}\ }\textbf {\bibinfo
  {volume} {195}},\ \bibinfo {pages} {L261} (\bibinfo {year}
  {1999})}\BibitemShut {NoStop}%
\bibitem [{\citenamefont {Berger}(1996)}]{Berger1996}%
  \BibitemOpen
  \bibfield  {author} {\bibinfo {author} {\bibfnamefont {L.}~\bibnamefont
  {Berger}},\ }\href {\doibase http://dx.doi.org/10.1103/PhysRevB.54.9353}
  {\bibfield  {journal} {\bibinfo  {journal} {Phys. Rev. B}\ }\textbf {\bibinfo
  {volume} {54}},\ \bibinfo {pages} {9353} (\bibinfo {year}
  {1996})}\BibitemShut {NoStop}%
\bibitem [{\citenamefont {Berger}(1999)}]{Berger1999}%
  \BibitemOpen
  \bibfield  {author} {\bibinfo {author} {\bibfnamefont {L.}~\bibnamefont
  {Berger}},\ }\href {\doibase 10.1103/PhysRevB.59.11465} {\bibfield  {journal}
  {\bibinfo  {journal} {Phys. Rev. B}\ }\textbf {\bibinfo {volume} {59}},\
  \bibinfo {pages} {11465} (\bibinfo {year} {1999})}\BibitemShut {NoStop}%
\bibitem [{\citenamefont {Chernyshov}\ \emph {et~al.}(2008)\citenamefont
  {Chernyshov}, \citenamefont {Overby}, \citenamefont {Liu}, \citenamefont
  {Furdyna}, \citenamefont {Lyanda-Geller},\ and\ \citenamefont
  {Rokhinson}}]{Chernyshov2008}%
  \BibitemOpen
  \bibfield  {author} {\bibinfo {author} {\bibfnamefont {A.}~\bibnamefont
  {Chernyshov}}, \bibinfo {author} {\bibfnamefont {M.}~\bibnamefont {Overby}},
  \bibinfo {author} {\bibfnamefont {X.}~\bibnamefont {Liu}}, \bibinfo {author}
  {\bibfnamefont {J.~K.}\ \bibnamefont {Furdyna}}, \bibinfo {author}
  {\bibfnamefont {Y.}~\bibnamefont {Lyanda-Geller}}, \ and\ \bibinfo {author}
  {\bibfnamefont {L.~P.}\ \bibnamefont {Rokhinson}},\ }\href {\doibase
  10.1038/nphys1362} {\bibfield  {journal} {\bibinfo  {journal} {Nat. Phys.}\
  }\textbf {\bibinfo {volume} {5}},\ \bibinfo {pages} {656} (\bibinfo {year}
  {2008})}\BibitemShut {NoStop}%
\bibitem [{\citenamefont {Manchon}\ and\ \citenamefont
  {Zhang}(2008)}]{Manchon2008}%
  \BibitemOpen
  \bibfield  {author} {\bibinfo {author} {\bibfnamefont {A.}~\bibnamefont
  {Manchon}}\ and\ \bibinfo {author} {\bibfnamefont {S.}~\bibnamefont
  {Zhang}},\ }\href {\doibase 10.1103/PhysRevB.78.212405} {\bibfield  {journal}
  {\bibinfo  {journal} {Phys. Rev. B}\ }\textbf {\bibinfo {volume} {78}},\
  \bibinfo {pages} {212405} (\bibinfo {year} {2008})}\BibitemShut {NoStop}%
\bibitem [{\citenamefont {Garate}\ and\ \citenamefont
  {MacDonald}(2009)}]{Garate2009}%
  \BibitemOpen
  \bibfield  {author} {\bibinfo {author} {\bibfnamefont {I.}~\bibnamefont
  {Garate}}\ and\ \bibinfo {author} {\bibfnamefont {A.~H.}\ \bibnamefont
  {MacDonald}},\ }\href {\doibase 10.1103/PhysRevB.80.134403} {\bibfield
  {journal} {\bibinfo  {journal} {Phys. Rev. B}\ }\textbf {\bibinfo {volume}
  {80}},\ \bibinfo {pages} {134403} (\bibinfo {year} {2009})}\BibitemShut
  {NoStop}%
\bibitem [{\citenamefont {Miron}\ \emph {et~al.}(2010)\citenamefont {Miron},
  \citenamefont {Gaudin}, \citenamefont {Auffret}, \citenamefont {Rodmacq},
  \citenamefont {Schuhl}, \citenamefont {Pizzini}, \citenamefont {Vogel},\ and\
  \citenamefont {Gambardella}}]{Miron2010}%
  \BibitemOpen
  \bibfield  {author} {\bibinfo {author} {\bibfnamefont {I.~M.}\ \bibnamefont
  {Miron}}, \bibinfo {author} {\bibfnamefont {G.}~\bibnamefont {Gaudin}},
  \bibinfo {author} {\bibfnamefont {S.}~\bibnamefont {Auffret}}, \bibinfo
  {author} {\bibfnamefont {B.}~\bibnamefont {Rodmacq}}, \bibinfo {author}
  {\bibfnamefont {A.}~\bibnamefont {Schuhl}}, \bibinfo {author} {\bibfnamefont
  {S.}~\bibnamefont {Pizzini}}, \bibinfo {author} {\bibfnamefont
  {J.}~\bibnamefont {Vogel}}, \ and\ \bibinfo {author} {\bibfnamefont
  {P.}~\bibnamefont {Gambardella}},\ }\href {\doibase 10.1038/nmat2613}
  {\bibfield  {journal} {\bibinfo  {journal} {Nat. Mater.}\ }\textbf {\bibinfo
  {volume} {9}},\ \bibinfo {pages} {230} (\bibinfo {year} {2010})}\BibitemShut
  {NoStop}%
\bibitem [{\citenamefont {Liu}\ \emph {et~al.}(2012)\citenamefont {Liu},
  \citenamefont {Lee}, \citenamefont {Gudmundsen}, \citenamefont {Ralph},\ and\
  \citenamefont {Buhrman}}]{Liu2012}%
  \BibitemOpen
  \bibfield  {author} {\bibinfo {author} {\bibfnamefont {L.}~\bibnamefont
  {Liu}}, \bibinfo {author} {\bibfnamefont {O.~J.}\ \bibnamefont {Lee}},
  \bibinfo {author} {\bibfnamefont {T.~J.}\ \bibnamefont {Gudmundsen}},
  \bibinfo {author} {\bibfnamefont {D.~C.}\ \bibnamefont {Ralph}}, \ and\
  \bibinfo {author} {\bibfnamefont {R.~A.}\ \bibnamefont {Buhrman}},\ }\href
  {\doibase 10.1103/PhysRevLett.109.096602} {\bibfield  {journal} {\bibinfo
  {journal} {Phys. Rev. Lett.}\ }\textbf {\bibinfo {volume} {109}},\ \bibinfo
  {pages} {096602} (\bibinfo {year} {2012})}\BibitemShut {NoStop}%
\bibitem [{\citenamefont {Khvalkovskiy}\ \emph {et~al.}(2013)\citenamefont
  {Khvalkovskiy}, \citenamefont {Cros}, \citenamefont {Apalkov}, \citenamefont
  {Nikitin}, \citenamefont {Krounbi}, \citenamefont {Zvezdin}, \citenamefont
  {Anane}, \citenamefont {Grollier},\ and\ \citenamefont
  {Fert}}]{Khvalkovskiy2013}%
  \BibitemOpen
  \bibfield  {author} {\bibinfo {author} {\bibfnamefont {A.~V.}\ \bibnamefont
  {Khvalkovskiy}}, \bibinfo {author} {\bibfnamefont {V.}~\bibnamefont {Cros}},
  \bibinfo {author} {\bibfnamefont {D.}~\bibnamefont {Apalkov}}, \bibinfo
  {author} {\bibfnamefont {V.}~\bibnamefont {Nikitin}}, \bibinfo {author}
  {\bibfnamefont {M.}~\bibnamefont {Krounbi}}, \bibinfo {author} {\bibfnamefont
  {K.~A.}\ \bibnamefont {Zvezdin}}, \bibinfo {author} {\bibfnamefont
  {A.}~\bibnamefont {Anane}}, \bibinfo {author} {\bibfnamefont
  {J.}~\bibnamefont {Grollier}}, \ and\ \bibinfo {author} {\bibfnamefont
  {A.}~\bibnamefont {Fert}},\ }\href {\doibase 10.1103/PhysRevB.87.020402}
  {\bibfield  {journal} {\bibinfo  {journal} {Phys. Rev. B}\ }\textbf {\bibinfo
  {volume} {87}},\ \bibinfo {pages} {020402} (\bibinfo {year}
  {2013})}\BibitemShut {NoStop}%
\bibitem [{\citenamefont {Hals}\ and\ \citenamefont
  {Brataas}(2013)}]{Hals2013}%
  \BibitemOpen
  \bibfield  {author} {\bibinfo {author} {\bibfnamefont {K.~M.~D.}\
  \bibnamefont {Hals}}\ and\ \bibinfo {author} {\bibfnamefont {A.}~\bibnamefont
  {Brataas}},\ }\href {\doibase 10.1103/PhysRevB.87.174409} {\bibfield
  {journal} {\bibinfo  {journal} {Phys. Rev. B}\ }\textbf {\bibinfo {volume}
  {87}},\ \bibinfo {pages} {174409} (\bibinfo {year} {2013})}\BibitemShut
  {NoStop}%
\bibitem [{\citenamefont {Hals}\ and\ \citenamefont
  {Brataas}(2014)}]{Hals2014}%
  \BibitemOpen
  \bibfield  {author} {\bibinfo {author} {\bibfnamefont {K.~M.~D.}\
  \bibnamefont {Hals}}\ and\ \bibinfo {author} {\bibfnamefont {A.}~\bibnamefont
  {Brataas}},\ }\href {\doibase 10.1103/PhysRevB.89.064426} {\bibfield
  {journal} {\bibinfo  {journal} {Phys. Rev. B}\ }\textbf {\bibinfo {volume}
  {89}},\ \bibinfo {pages} {064426} (\bibinfo {year} {2014})}\BibitemShut
  {NoStop}%
\bibitem [{\citenamefont {Bergeret}\ \emph {et~al.}(2001)\citenamefont
  {Bergeret}, \citenamefont {Volkov},\ and\ \citenamefont
  {Efetov}}]{Bergeret2001}%
  \BibitemOpen
  \bibfield  {author} {\bibinfo {author} {\bibfnamefont {F.~S.}\ \bibnamefont
  {Bergeret}}, \bibinfo {author} {\bibfnamefont {A.~F.}\ \bibnamefont
  {Volkov}}, \ and\ \bibinfo {author} {\bibfnamefont {K.~B.}\ \bibnamefont
  {Efetov}},\ }\href {\doibase 10.1103/PhysRevLett.86.4096} {\bibfield
  {journal} {\bibinfo  {journal} {Phys. Rev. Lett.}\ }\textbf {\bibinfo
  {volume} {86}},\ \bibinfo {pages} {4096} (\bibinfo {year}
  {2001})}\BibitemShut {NoStop}%
\bibitem [{\citenamefont {Waintal}\ and\ \citenamefont
  {Brouwer}(2002)}]{Waintal2002}%
  \BibitemOpen
  \bibfield  {author} {\bibinfo {author} {\bibfnamefont {X.}~\bibnamefont
  {Waintal}}\ and\ \bibinfo {author} {\bibfnamefont {P.~W.}\ \bibnamefont
  {Brouwer}},\ }\href {\doibase 10.1103/PhysRevB.65.054407} {\bibfield
  {journal} {\bibinfo  {journal} {Phys. Rev. B}\ }\textbf {\bibinfo {volume}
  {65}},\ \bibinfo {pages} {054407} (\bibinfo {year} {2002})}\BibitemShut
  {NoStop}%
\bibitem [{\citenamefont {Bergeret}\ \emph {et~al.}(2005)\citenamefont
  {Bergeret}, \citenamefont {Volkov},\ and\ \citenamefont
  {Efetov}}]{Bergeret2005}%
  \BibitemOpen
  \bibfield  {author} {\bibinfo {author} {\bibfnamefont {F.~S.}\ \bibnamefont
  {Bergeret}}, \bibinfo {author} {\bibfnamefont {A.~F.}\ \bibnamefont
  {Volkov}}, \ and\ \bibinfo {author} {\bibfnamefont {K.~B.}\ \bibnamefont
  {Efetov}},\ }\href {\doibase 10.1103/RevModPhys.77.1321} {\bibfield
  {journal} {\bibinfo  {journal} {Rev. Mod. Phys.}\ }\textbf {\bibinfo {volume}
  {77}},\ \bibinfo {pages} {1321} (\bibinfo {year} {2005})}\BibitemShut
  {NoStop}%
\bibitem [{\citenamefont {Keizer}\ \emph {et~al.}(2006)\citenamefont {Keizer},
  \citenamefont {Goennenwein}, \citenamefont {Klapwijk}, \citenamefont {Miao},
  \citenamefont {Xiao},\ and\ \citenamefont {Gupta}}]{Keizer2006}%
  \BibitemOpen
  \bibfield  {author} {\bibinfo {author} {\bibfnamefont {R.~S.}\ \bibnamefont
  {Keizer}}, \bibinfo {author} {\bibfnamefont {S.~T.~B.}\ \bibnamefont
  {Goennenwein}}, \bibinfo {author} {\bibfnamefont {T.~M.}\ \bibnamefont
  {Klapwijk}}, \bibinfo {author} {\bibfnamefont {G.}~\bibnamefont {Miao}},
  \bibinfo {author} {\bibfnamefont {G.}~\bibnamefont {Xiao}}, \ and\ \bibinfo
  {author} {\bibfnamefont {A.}~\bibnamefont {Gupta}},\ }\href {\doibase
  10.1038/nature04499} {\bibfield  {journal} {\bibinfo  {journal} {Nature (London)}\
  }\textbf {\bibinfo {volume} {439}},\ \bibinfo {pages} {825} (\bibinfo {year}
  {2006})}\BibitemShut {NoStop}%
\bibitem [{\citenamefont {Eschrig}\ and\ \citenamefont
  {L{\"{o}}fwander}(2008)}]{Eschrig2008}%
  \BibitemOpen
  \bibfield  {author} {\bibinfo {author} {\bibfnamefont {M.}~\bibnamefont
  {Eschrig}}\ and\ \bibinfo {author} {\bibfnamefont {T.}~\bibnamefont
  {L{\"{o}}fwander}},\ }\href {\doibase 10.1038/nphys831} {\bibfield  {journal}
  {\bibinfo  {journal} {Nat. Phys.}\ }\textbf {\bibinfo {volume} {4}},\
  \bibinfo {pages} {138} (\bibinfo {year} {2008})}\BibitemShut {NoStop}%
\bibitem [{\citenamefont {Zhao}\ and\ \citenamefont {Sauls}(2008)}]{Zhao2008}%
  \BibitemOpen
  \bibfield  {author} {\bibinfo {author} {\bibfnamefont {E.}~\bibnamefont
  {Zhao}}\ and\ \bibinfo {author} {\bibfnamefont {J.~A.}\ \bibnamefont
  {Sauls}},\ }\href {\doibase 10.1103/PhysRevB.78.174511} {\bibfield  {journal}
  {\bibinfo  {journal} {Phys. Rev. B}\ }\textbf {\bibinfo {volume} {78}},\
  \bibinfo {pages} {174511} (\bibinfo {year} {2008})}\BibitemShut {NoStop}%
\bibitem [{\citenamefont {Braude}\ and\ \citenamefont
  {Blanter}(2008)}]{Braude2008}%
  \BibitemOpen
  \bibfield  {author} {\bibinfo {author} {\bibfnamefont {V.}~\bibnamefont
  {Braude}}\ and\ \bibinfo {author} {\bibfnamefont {Y.~M.}\ \bibnamefont
  {Blanter}},\ }\href {http://prl.aps.org/pdf/PRL/v100/i20/e207001} {\bibfield
  {journal} {\bibinfo  {journal} {Phys. Rev. Lett.}\ }\textbf {\bibinfo
  {volume} {100}},\ \bibinfo {pages} {207001} (\bibinfo {year}
  {2008})}\BibitemShut {NoStop}%
\bibitem [{\citenamefont {Konschelle}\ and\ \citenamefont
  {Buzdin}(2009)}]{Konschelle2009}%
  \BibitemOpen
  \bibfield  {author} {\bibinfo {author} {\bibfnamefont {F.}~\bibnamefont
  {Konschelle}}\ and\ \bibinfo {author} {\bibfnamefont {A.}~\bibnamefont
  {Buzdin}},\ }\href {\doibase 10.1103/PhysRevLett.102.017001} {\bibfield
  {journal} {\bibinfo  {journal} {Phys. Rev. Lett.}\ }\textbf {\bibinfo
  {volume} {102}},\ \bibinfo {pages} {017001} (\bibinfo {year}
  {2009})}\BibitemShut {NoStop}%
\bibitem [{\citenamefont {Eschrig}(2011)}]{Eschrig2011}%
  \BibitemOpen
  \bibfield  {author} {\bibinfo {author} {\bibfnamefont {M.}~\bibnamefont
  {Eschrig}},\ }\href {\doibase 10.1063/1.3541944} {\bibfield  {journal}
  {\bibinfo  {journal} {Phys. Today}\ }\textbf {\bibinfo {volume} {64(1)}},\
  \bibinfo {pages} {43} (\bibinfo {year} {2011})}, \BibitemShut {NoStop}%
\bibitem [{\citenamefont {Linder}\ and\ \citenamefont
  {Yokoyama}(2011)}]{Linder2011}%
  \BibitemOpen
  \bibfield  {author} {\bibinfo {author} {\bibfnamefont {J.}~\bibnamefont
  {Linder}}\ and\ \bibinfo {author} {\bibfnamefont {T.}~\bibnamefont
  {Yokoyama}},\ }\href {\doibase 10.1103/PhysRevB.83.012501} {\bibfield
  {journal} {\bibinfo  {journal} {Phys. Rev. B}\ }\textbf {\bibinfo {volume}
  {83}},\ \bibinfo {pages} {012501} (\bibinfo {year} {2011})}\BibitemShut
  {NoStop}%
\bibitem [{\citenamefont {Sacramento}\ \emph {et~al.}(2011)\citenamefont
  {Sacramento}, \citenamefont {{Fernandes Silva}}, \citenamefont {Nunes},
  \citenamefont {Ara{\'{u}}jo},\ and\ \citenamefont
  {Vieira}}]{Sacramento2011a}%
  \BibitemOpen
  \bibfield  {author} {\bibinfo {author} {\bibfnamefont {P.}~\bibnamefont
  {Sacramento}}, \bibinfo {author} {\bibfnamefont {L.}~\bibnamefont {{Fernandes
  Silva}}}, \bibinfo {author} {\bibfnamefont {G.}~\bibnamefont {Nunes}},
  \bibinfo {author} {\bibfnamefont {M.}~\bibnamefont {Ara{\'{u}}jo}}, \ and\
  \bibinfo {author} {\bibfnamefont {V.}~\bibnamefont {Vieira}},\ }\href
  {\doibase 10.1103/PhysRevB.83.054403} {\bibfield  {journal} {\bibinfo
  {journal} {Phys. Rev. B}\ }\textbf {\bibinfo {volume} {83}},\ \bibinfo
  {pages} {054403} (\bibinfo {year} {2011})}\BibitemShut {NoStop}%
\bibitem [{\citenamefont {Linder}\ \emph {et~al.}(2012)\citenamefont {Linder},
  \citenamefont {Brataas}, \citenamefont {Shomali},\ and\ \citenamefont
  {Zareyan}}]{Linder2012}%
  \BibitemOpen
  \bibfield  {author} {\bibinfo {author} {\bibfnamefont {J.}~\bibnamefont
  {Linder}}, \bibinfo {author} {\bibfnamefont {A.}~\bibnamefont {Brataas}},
  \bibinfo {author} {\bibfnamefont {Z.}~\bibnamefont {Shomali}}, \ and\
  \bibinfo {author} {\bibfnamefont {M.}~\bibnamefont {Zareyan}},\ }\href
  {\doibase 10.1103/PhysRevLett.109.237206} {\bibfield  {journal} {\bibinfo
  {journal} {Phys. Rev. Lett.}\ }\textbf {\bibinfo {volume} {109}},\ \bibinfo
  {pages} {237206} (\bibinfo {year} {2012})}\BibitemShut {NoStop}%
\bibitem [{\citenamefont {Bergeret}\ and\ \citenamefont
  {Tokatly}(2013)}]{Bergeret2013}%
  \BibitemOpen
  \bibfield  {author} {\bibinfo {author} {\bibfnamefont {F.~S.}\ \bibnamefont
  {Bergeret}}\ and\ \bibinfo {author} {\bibfnamefont {I.~V.}\ \bibnamefont
  {Tokatly}},\ }\href {\doibase 10.1103/PhysRevLett.110.117003} {\bibfield
  {journal} {\bibinfo  {journal} {Phys. Rev. Lett.}\ }\textbf {\bibinfo
  {volume} {110}},\ \bibinfo {pages} {117003} (\bibinfo {year}
  {2013})}\BibitemShut {NoStop}%
\bibitem [{\citenamefont {Bergeret}\ and\ \citenamefont
  {Tokatly}(2014)}]{Bergeret2014}%
  \BibitemOpen
  \bibfield  {author} {\bibinfo {author} {\bibfnamefont {F.~S.}\ \bibnamefont
  {Bergeret}}\ and\ \bibinfo {author} {\bibfnamefont {I.~V.}\ \bibnamefont
  {Tokatly}},\ }\href {\doibase 10.1103/PhysRevB.89.134517} {\bibfield
  {journal} {\bibinfo  {journal} {Phys. Rev. B}\ }\textbf {\bibinfo {volume}
  {89}},\ \bibinfo {pages} {134517} (\bibinfo {year} {2014})}\BibitemShut
  {NoStop}%
\bibitem [{\citenamefont {Kulagina}\ and\ \citenamefont
  {Linder}(2014)}]{Kulagina2014}%
  \BibitemOpen
  \bibfield  {author} {\bibinfo {author} {\bibfnamefont {I.}~\bibnamefont
  {Kulagina}}\ and\ \bibinfo {author} {\bibfnamefont {J.}~\bibnamefont
  {Linder}},\ }\href {\doibase 10.1103/PhysRevB.90.054504} {\bibfield
  {journal} {\bibinfo  {journal} {Phys. Rev. B}\ }\textbf {\bibinfo {volume}
  {90}},\ \bibinfo {pages} {054504} (\bibinfo {year} {2014})}\BibitemShut
  {NoStop}%
\bibitem [{\citenamefont {Halterman}\ \emph {et~al.}(2015)\citenamefont
  {Halterman}, \citenamefont {Valls},\ and\ \citenamefont
  {Wu}}]{Halterman2015a}%
  \BibitemOpen
  \bibfield  {author} {\bibinfo {author} {\bibfnamefont {K.}~\bibnamefont
  {Halterman}}, \bibinfo {author} {\bibfnamefont {O.~T.}\ \bibnamefont
  {Valls}}, \ and\ \bibinfo {author} {\bibfnamefont {C.-T.}\ \bibnamefont
  {Wu}},\ }\href {\doibase 10.1103/PhysRevB.92.174516} {\bibfield  {journal}
  {\bibinfo  {journal} {Phys. Rev. B}\ }\textbf {\bibinfo {volume} {92}},\
  \bibinfo {pages} {174516} (\bibinfo {year} {2015})}\BibitemShut {NoStop}%
\bibitem [{\citenamefont {Linder}\ and\ \citenamefont
  {Robinson}(2015)}]{Linder2015}%
  \BibitemOpen
  \bibfield  {author} {\bibinfo {author} {\bibfnamefont {J.}~\bibnamefont
  {Linder}}\ and\ \bibinfo {author} {\bibfnamefont {J.~W.~A.}\ \bibnamefont
  {Robinson}},\ }\href {\doibase 10.1038/nphys3242} {\bibfield  {journal}
  {\bibinfo  {journal} {Nat. Phys.}\ }\textbf {\bibinfo {volume} {11}},\
  \bibinfo {pages} {307} (\bibinfo {year} {2015})}\BibitemShut {NoStop}%
\bibitem [{\citenamefont {Eschrig}(2015)}]{Eschrig2015a}%
  \BibitemOpen
  \bibfield  {author} {\bibinfo {author} {\bibfnamefont {M.}~\bibnamefont
  {Eschrig}},\ }\href {\doibase 10.1088/0034-4885/78/10/104501} {\bibfield
  {journal} {\bibinfo  {journal} {Rep. Prog. Phys.}\ }\textbf {\bibinfo
  {volume} {78}},\ \bibinfo {pages} {104501} (\bibinfo {year}
  {2015})}\BibitemShut {NoStop}%
\bibitem [{\citenamefont {Jacobsen}\ \emph {et~al.}(2015)\citenamefont
  {Jacobsen}, \citenamefont {Ouassou},\ and\ \citenamefont
  {Linder}}]{Jacobsen2015}%
  \BibitemOpen
  \bibfield  {author} {\bibinfo {author} {\bibfnamefont {S.~H.}\ \bibnamefont
  {Jacobsen}}, \bibinfo {author} {\bibfnamefont {J.~A.}\ \bibnamefont
  {Ouassou}}, \ and\ \bibinfo {author} {\bibfnamefont {J.}~\bibnamefont
  {Linder}},\ }\href {\doibase 10.1103/PhysRevB.92.024510} {\bibfield
  {journal} {\bibinfo  {journal} {Phys. Rev. B}\ }\textbf {\bibinfo {volume}
  {92}},\ \bibinfo {pages} {024510} (\bibinfo {year} {2015})}\BibitemShut
  {NoStop}%
\bibitem [{\citenamefont {Yokoyama}\ and\ \citenamefont
  {Linder}(2015)}]{Yokoyama2015a}%
  \BibitemOpen
  \bibfield  {author} {\bibinfo {author} {\bibfnamefont {T.}~\bibnamefont
  {Yokoyama}}\ and\ \bibinfo {author} {\bibfnamefont {J.}~\bibnamefont
  {Linder}},\ }\href {\doibase 10.1103/PhysRevB.92.060503} {\bibfield
  {journal} {\bibinfo  {journal} {Phys. Rev. B}\ }\textbf {\bibinfo {volume}
  {92}},\ \bibinfo {pages} {060503} (\bibinfo {year} {2015})}\BibitemShut
  {NoStop}%
\bibitem [{\citenamefont {Hals}(2016)}]{Hals2016}%
  \BibitemOpen
  \bibfield  {author} {\bibinfo {author} {\bibfnamefont {K.~M.~D.}\
  \bibnamefont {Hals}},\ }\href {\doibase 10.1103/PhysRevB.93.115431}
  {\bibfield  {journal} {\bibinfo  {journal} {Phys. Rev. B}\ }\textbf {\bibinfo
  {volume} {93}},\ \bibinfo {pages} {115431} (\bibinfo {year}
  {2016})}\BibitemShut {NoStop}%
\bibitem [{\citenamefont {Fujimoto}(2008)}]{Fujimoto2008}%
  \BibitemOpen
  \bibfield  {author} {\bibinfo {author} {\bibfnamefont {S.}~\bibnamefont
  {Fujimoto}},\ }\href {\doibase 10.1103/PhysRevB.77.220501} {\bibfield
  {journal} {\bibinfo  {journal} {Phys. Rev. B}\ }\textbf {\bibinfo {volume}
  {77}},\ \bibinfo {pages} {220501} (\bibinfo {year} {2008})}\BibitemShut
  {NoStop}%
\bibitem [{\citenamefont {Sato}\ \emph {et~al.}(2009)\citenamefont {Sato},
  \citenamefont {Takahashi},\ and\ \citenamefont {Fujimoto}}]{Sato2009}%
  \BibitemOpen
  \bibfield  {author} {\bibinfo {author} {\bibfnamefont {M.}~\bibnamefont
  {Sato}}, \bibinfo {author} {\bibfnamefont {Y.}~\bibnamefont {Takahashi}}, \
  and\ \bibinfo {author} {\bibfnamefont {S.}~\bibnamefont {Fujimoto}},\ }\href
  {\doibase 10.1103/PhysRevLett.103.020401} {\bibfield  {journal} {\bibinfo
  {journal} {Phys. Rev. Lett.}\ }\textbf {\bibinfo {volume} {103}},\ \bibinfo
  {pages} {020401} (\bibinfo {year} {2009})}\BibitemShut {NoStop}%
\bibitem [{\citenamefont {Sau}\ \emph {et~al.}(2010)\citenamefont {Sau},
  \citenamefont {Lutchyn}, \citenamefont {Tewari},\ and\ \citenamefont {{Das
  Sarma}}}]{Sau2010}%
  \BibitemOpen
  \bibfield  {author} {\bibinfo {author} {\bibfnamefont {J.~D.}\ \bibnamefont
  {Sau}}, \bibinfo {author} {\bibfnamefont {R.~M.}\ \bibnamefont {Lutchyn}},
  \bibinfo {author} {\bibfnamefont {S.}~\bibnamefont {Tewari}}, \ and\ \bibinfo
  {author} {\bibfnamefont {S.}~\bibnamefont {{Das Sarma}}},\ }\href {\doibase
  10.1103/PhysRevLett.104.040502} {\bibfield  {journal} {\bibinfo  {journal}
  {Phys. Rev. Lett.}\ }\textbf {\bibinfo {volume} {104}},\ \bibinfo {pages}
  {040502} (\bibinfo {year} {2010})}\BibitemShut {NoStop}%
\bibitem [{\citenamefont {Pershoguba}\ \emph {et~al.}(2015)\citenamefont
  {Pershoguba}, \citenamefont {Bj\"{o}rnson}, \citenamefont
  {Black-Schaffer},\ and\ \citenamefont {Balatsky}}]{Pershoguba2015a}%
  \BibitemOpen
  \bibfield  {author} {\bibinfo {author} {\bibfnamefont {S.~S.}\ \bibnamefont
  {Pershoguba}}, \bibinfo {author} {\bibfnamefont {K.}~\bibnamefont
  {Bj\"{o}rnson}}, \bibinfo {author} {\bibfnamefont {A.~M.}\
  \bibnamefont {Black-Schaffer}}, \ and\ \bibinfo {author} {\bibfnamefont
  {A.~V.}\ \bibnamefont {Balatsky}},\ }\href {\doibase
  10.1103/PhysRevLett.115.116602} {\bibfield  {journal} {\bibinfo  {journal}
  {Phys. Rev. Lett.}\ }\textbf {\bibinfo {volume} {115}},\ \bibinfo {pages}
  {116602} (\bibinfo {year} {2015})}\BibitemShut {NoStop}%
\bibitem [{\citenamefont {Bj\"{o}rnson}\ \emph
  {et~al.}(2015)\citenamefont {Bj\"{o}rnson}, \citenamefont
  {Pershoguba}, \citenamefont {Balatsky},\ and\ \citenamefont
  {Black-Schaffer}}]{Bjornson2015}%
  \BibitemOpen
  \bibfield  {author} {\bibinfo {author} {\bibfnamefont {K.}~\bibnamefont
  {Bj\"ornson}}, \bibinfo {author} {\bibfnamefont {S.~S.}\
  \bibnamefont {Pershoguba}}, \bibinfo {author} {\bibfnamefont {A.~V.}\
  \bibnamefont {Balatsky}}, \ and\ \bibinfo {author} {\bibfnamefont {A.~M.}\
  \bibnamefont {Black-Schaffer}},\ }\href {\doibase 10.1103/PhysRevB.92.214501}
  {\bibfield  {journal} {\bibinfo  {journal} {Phys. Rev. B}\ }\textbf {\bibinfo
  {volume} {92}},\ \bibinfo {pages} {214501} (\bibinfo {year}
  {2015})}\BibitemShut {NoStop}%
\bibitem [{\citenamefont {Nakosai}\ \emph {et~al.}(2013)\citenamefont
  {Nakosai}, \citenamefont {Tanaka},\ and\ \citenamefont
  {Nagaosa}}]{Nakosai2013}%
  \BibitemOpen
  \bibfield  {author} {\bibinfo {author} {\bibfnamefont {S.}~\bibnamefont
  {Nakosai}}, \bibinfo {author} {\bibfnamefont {Y.}~\bibnamefont {Tanaka}}, \
  and\ \bibinfo {author} {\bibfnamefont {N.}~\bibnamefont {Nagaosa}},\ }\href
  {\doibase 10.1103/PhysRevB.88.180503} {\bibfield  {journal} {\bibinfo
  {journal} {Phys. Rev. B}\ }\textbf {\bibinfo {volume} {88}},\ \bibinfo
  {pages} {180503} (\bibinfo {year} {2013})}\BibitemShut {NoStop}%
\bibitem [{\citenamefont {Klinovaja}\ \emph {et~al.}(2013)\citenamefont
  {Klinovaja}, \citenamefont {Stano}, \citenamefont {Yazdani},\ and\
  \citenamefont {Loss}}]{Klinovaja2013}%
  \BibitemOpen
  \bibfield  {author} {\bibinfo {author} {\bibfnamefont {J.}~\bibnamefont
  {Klinovaja}}, \bibinfo {author} {\bibfnamefont {P.}~\bibnamefont {Stano}},
  \bibinfo {author} {\bibfnamefont {A.}~\bibnamefont {Yazdani}}, \ and\
  \bibinfo {author} {\bibfnamefont {D.}~\bibnamefont {Loss}},\ }\href {\doibase
  10.1103/PhysRevLett.111.186805} {\bibfield  {journal} {\bibinfo  {journal}
  {Phys. Rev. Lett.}\ }\textbf {\bibinfo {volume} {111}},\ \bibinfo {pages}
  {186805} (\bibinfo {year} {2013})}\BibitemShut {NoStop}%
\bibitem [{\citenamefont {Chen}\ and\ \citenamefont
  {Schnyder}(2015)}]{Chen2015a}%
  \BibitemOpen
  \bibfield  {author} {\bibinfo {author} {\bibfnamefont {W.}~\bibnamefont
  {Chen}}\ and\ \bibinfo {author} {\bibfnamefont {A.~P.}\ \bibnamefont
  {Schnyder}},\ }\href {\doibase 10.1103/PhysRevB.92.214502} {\bibfield
  {journal} {\bibinfo  {journal} {Phys. Rev. B}\ }\textbf {\bibinfo {volume}
  {92}},\ \bibinfo {pages} {214502} (\bibinfo {year} {2015})}\BibitemShut
  {NoStop}%
\bibitem [{\citenamefont {Li}\ \emph {et~al.}(2016)\citenamefont {Li},
  \citenamefont {Neupert}, \citenamefont {Wang}, \citenamefont {MacDonald},
  \citenamefont {Yazdani},\ and\ \citenamefont {Bernevig}}]{Li2016}%
  \BibitemOpen
  \bibfield  {author} {\bibinfo {author} {\bibfnamefont {J.}~\bibnamefont
  {Li}}, \bibinfo {author} {\bibfnamefont {T.}~\bibnamefont {Neupert}},
  \bibinfo {author} {\bibfnamefont {Z.}~\bibnamefont {Wang}}, \bibinfo {author}
  {\bibfnamefont {A.~H.}\ \bibnamefont {MacDonald}}, \bibinfo {author}
  {\bibfnamefont {A.}~\bibnamefont {Yazdani}}, \ and\ \bibinfo {author}
  {\bibfnamefont {B.~A.}\ \bibnamefont {Bernevig}},\ }\href {\doibase
  10.1038/ncomms12297} {\bibfield  {journal} {\bibinfo  {journal} {Nat.
  Commun.}\ }\textbf {\bibinfo {volume} {7}},\ \bibinfo {pages} {12297}
  (\bibinfo {year} {2016})}\BibitemShut {NoStop}%
\bibitem [{\citenamefont {Volovik}(1998)}]{Volovik1998}%
  \BibitemOpen
  \bibfield  {author} {\bibinfo {author} {\bibfnamefont {G.}~\bibnamefont
  {Volovik}},\ }\href {\doibase 10.1016/S0921-4526(98)00456-6} {\bibfield
  {journal} {\bibinfo  {journal} {Phys. B Condens. Matter}\ }\textbf {\bibinfo
  {volume} {255}},\ \bibinfo {pages} {86} (\bibinfo {year} {1998})}\BibitemShut
  {NoStop}%
\bibitem [{\citenamefont {Bevan}\ \emph {et~al.}(1997)\citenamefont {Bevan},
  \citenamefont {Manninen}, \citenamefont {Cook}, \citenamefont {Hook},
  \citenamefont {Hall}, \citenamefont {Vachaspati},\ and\ \citenamefont
  {Volovik}}]{Bevan1997}%
  \BibitemOpen
  \bibfield  {author} {\bibinfo {author} {\bibfnamefont {T.~D.~C.}\
  \bibnamefont {Bevan}}, \bibinfo {author} {\bibfnamefont {A.~J.}\ \bibnamefont
  {Manninen}}, \bibinfo {author} {\bibfnamefont {J.~B.}\ \bibnamefont {Cook}},
  \bibinfo {author} {\bibfnamefont {J.~R.}\ \bibnamefont {Hook}}, \bibinfo
  {author} {\bibfnamefont {H.~E.}\ \bibnamefont {Hall}}, \bibinfo {author}
  {\bibfnamefont {T.}~\bibnamefont {Vachaspati}}, \ and\ \bibinfo {author}
  {\bibfnamefont {G.~E.}\ \bibnamefont {Volovik}},\ }\href {\doibase
  10.1038/386689a0} {\bibfield  {journal} {\bibinfo  {journal} {Nature}\
  }\textbf {\bibinfo {volume} {386}},\ \bibinfo {pages} {689} (\bibinfo {year}
  {1997})}\BibitemShut {NoStop}%
\bibitem [{\citenamefont {Xu}\ \emph {et~al.}({\natexlab{a}})\citenamefont
  {Xu}, \citenamefont {Alidoust}, \citenamefont {Chang}, \citenamefont {Lu},
  \citenamefont {Singh}, \citenamefont {Belopolski}, \citenamefont {Sanchez},
  \citenamefont {Zhang}, \citenamefont {Bian}, \citenamefont {Zheng},
  \citenamefont {Husanu}, \citenamefont {Bian}, \citenamefont {Huang},
  \citenamefont {Hsu}, \citenamefont {Chang}, \citenamefont {Jeng},
  \citenamefont {Bansil}, \citenamefont {Strocov}, \citenamefont {Lin},
  \citenamefont {Jia},\ and\ \citenamefont {Hasan}}]{Xu2016}%
  \BibitemOpen
  \bibfield  {author} {\bibinfo {author} {\bibfnamefont {S.-Y.}\ \bibnamefont
  {Xu}}, \bibinfo {author} {\bibfnamefont {N.}~\bibnamefont {Alidoust}},
  \bibinfo {author} {\bibfnamefont {G.}~\bibnamefont {Chang}}, \bibinfo
  {author} {\bibfnamefont {H.}~\bibnamefont {Lu}}, \bibinfo {author}
  {\bibfnamefont {B.}~\bibnamefont {Singh}}, \bibinfo {author} {\bibfnamefont
  {I.}~\bibnamefont {Belopolski}}, \bibinfo {author} {\bibfnamefont
  {D.}~\bibnamefont {Sanchez}}, \bibinfo {author} {\bibfnamefont
  {X.}~\bibnamefont {Zhang}}, \bibinfo {author} {\bibfnamefont
  {G.}~\bibnamefont {Bian}}, \bibinfo {author} {\bibfnamefont {H.}~\bibnamefont
  {Zheng}}, \bibinfo {author} {\bibfnamefont {M.-A.}\ \bibnamefont {Husanu}},
  \bibinfo {author} {\bibfnamefont {Y.}~\bibnamefont {Bian}}, \bibinfo {author}
  {\bibfnamefont {S.-M.}\ \bibnamefont {Huang}}, \bibinfo {author}
  {\bibfnamefont {C.-H.}\ \bibnamefont {Hsu}}, \bibinfo {author} {\bibfnamefont
  {T.-R.}\ \bibnamefont {Chang}}, \bibinfo {author} {\bibfnamefont {H.-T.}\
  \bibnamefont {Jeng}}, \bibinfo {author} {\bibfnamefont {A.}~\bibnamefont
  {Bansil}}, \bibinfo {author} {\bibfnamefont {V.~N.}\ \bibnamefont {Strocov}},
  \bibinfo {author} {\bibfnamefont {H.}~\bibnamefont {Lin}}, \bibinfo {author}
  {\bibfnamefont {S.}~\bibnamefont {Jia}}, \ and\ \bibinfo {author}
  {\bibfnamefont {M.~Z.}\ \bibnamefont {Hasan}},\ }\href
  {http://arxiv.org/abs/1603.07318} {\ },\ \Eprint
  {http://arxiv.org/abs/1603.07318} {arXiv:1603.07318} \BibitemShut {NoStop}%
\bibitem [{\citenamefont {Huang}\ \emph {et~al.}()\citenamefont {Huang},
  \citenamefont {McCormick}, \citenamefont {Ochi}, \citenamefont {Zhao},
  \citenamefont {Suzuki}, \citenamefont {Arita}, \citenamefont {Wu},
  \citenamefont {Mou}, \citenamefont {Cao}, \citenamefont {Yan}, \citenamefont
  {Trivedi},\ and\ \citenamefont {Kaminski}}]{Huang2016}%
  \BibitemOpen
  \bibfield  {author} {\bibinfo {author} {\bibfnamefont {L.}~\bibnamefont
  {Huang}}, \bibinfo {author} {\bibfnamefont {T.~M.}\ \bibnamefont
  {McCormick}}, \bibinfo {author} {\bibfnamefont {M.}~\bibnamefont {Ochi}},
  \bibinfo {author} {\bibfnamefont {Z.}~\bibnamefont {Zhao}}, \bibinfo {author}
  {\bibfnamefont {M.}~\bibnamefont {Suzuki}}, \bibinfo {author} {\bibfnamefont
  {R.}~\bibnamefont {Arita}}, \bibinfo {author} {\bibfnamefont
  {Y.}~\bibnamefont {Wu}}, \bibinfo {author} {\bibfnamefont {D.}~\bibnamefont
  {Mou}}, \bibinfo {author} {\bibfnamefont {H.}~\bibnamefont {Cao}}, \bibinfo
  {author} {\bibfnamefont {J.}~\bibnamefont {Yan}}, \bibinfo {author}
  {\bibfnamefont {N.}~\bibnamefont {Trivedi}}, \ and\ \bibinfo {author}
  {\bibfnamefont {A.}~\bibnamefont {Kaminski}},\ }\href
  {http://arxiv.org/abs/1603.06482} {\ }\Eprint
  {http://arxiv.org/abs/1603.06482} {Nat. Mater. {\bf 15},1155 (2016) } \BibitemShut {NoStop}%
\bibitem [{\citenamefont {Xu}\ \emph {et~al.}({\natexlab{b}})\citenamefont
  {Xu}, \citenamefont {Wang}, \citenamefont {Weber}, \citenamefont {Magrez},
  \citenamefont {Bugnon}, \citenamefont {Berger}, \citenamefont {Matt},
  \citenamefont {Ma}, \citenamefont {Fu}, \citenamefont {Lv}, \citenamefont
  {Plumb}, \citenamefont {Radovic}, \citenamefont {Pomjakushina}, \citenamefont
  {Conder}, \citenamefont {Qian}, \citenamefont {Dil}, \citenamefont {Mesot},
  \citenamefont {Ding},\ and\ \citenamefont {Shi}}]{Xu2016a}%
  \BibitemOpen
  \bibfield  {author} {\bibinfo {author} {\bibfnamefont {N.}~\bibnamefont
  {Xu}}, \bibinfo {author} {\bibfnamefont {Z.~J.}\ \bibnamefont {Wang}},
  \bibinfo {author} {\bibfnamefont {A.~P.}\ \bibnamefont {Weber}}, \bibinfo
  {author} {\bibfnamefont {A.}~\bibnamefont {Magrez}}, \bibinfo {author}
  {\bibfnamefont {P.}~\bibnamefont {Bugnon}}, \bibinfo {author} {\bibfnamefont
  {H.}~\bibnamefont {Berger}}, \bibinfo {author} {\bibfnamefont {C.~E.}\
  \bibnamefont {Matt}}, \bibinfo {author} {\bibfnamefont {J.~Z.}\ \bibnamefont
  {Ma}}, \bibinfo {author} {\bibfnamefont {B.~B.}\ \bibnamefont {Fu}}, \bibinfo
  {author} {\bibfnamefont {B.~Q.}\ \bibnamefont {Lv}}, \bibinfo {author}
  {\bibfnamefont {N.~C.}\ \bibnamefont {Plumb}}, \bibinfo {author}
  {\bibfnamefont {M.}~\bibnamefont {Radovic}}, \bibinfo {author} {\bibfnamefont
  {E.}~\bibnamefont {Pomjakushina}}, \bibinfo {author} {\bibfnamefont
  {K.}~\bibnamefont {Conder}}, \bibinfo {author} {\bibfnamefont
  {T.}~\bibnamefont {Qian}}, \bibinfo {author} {\bibfnamefont {J.~H.}\
  \bibnamefont {Dil}}, \bibinfo {author} {\bibfnamefont {J.}~\bibnamefont
  {Mesot}}, \bibinfo {author} {\bibfnamefont {H.}~\bibnamefont {Ding}}, \ and\
  \bibinfo {author} {\bibfnamefont {M.}~\bibnamefont {Shi}},\ }\ \Eprint
  {http://arxiv.org/abs/1604.02116} {arXiv:1604.02116} \BibitemShut {NoStop}%
\bibitem [{\citenamefont {Deng}\ \emph {et~al.}()\citenamefont {Deng},
  \citenamefont {Wan}, \citenamefont {Deng}, \citenamefont {Zhang},
  \citenamefont {Ding}, \citenamefont {Wang}, \citenamefont {Yan},
  \citenamefont {Huang}, \citenamefont {Zhang}, \citenamefont {Xu},
  \citenamefont {Denlinger}, \citenamefont {Fedorov}, \citenamefont {Yang},
  \citenamefont {Duan}, \citenamefont {Yao}, \citenamefont {Wu}, \citenamefont
  {Fan}, \citenamefont {Zhang}, \citenamefont {Chen},\ and\ \citenamefont
  {Zhou}}]{Deng2016}%
  \BibitemOpen
  \bibfield  {author} {\bibinfo {author} {\bibfnamefont {K.}~\bibnamefont
  {Deng}}, \bibinfo {author} {\bibfnamefont {G.}~\bibnamefont {Wan}}, \bibinfo
  {author} {\bibfnamefont {P.}~\bibnamefont {Deng}}, \bibinfo {author}
  {\bibfnamefont {K.}~\bibnamefont {Zhang}}, \bibinfo {author} {\bibfnamefont
  {S.}~\bibnamefont {Ding}}, \bibinfo {author} {\bibfnamefont {E.}~\bibnamefont
  {Wang}}, \bibinfo {author} {\bibfnamefont {M.}~\bibnamefont {Yan}}, \bibinfo
  {author} {\bibfnamefont {H.}~\bibnamefont {Huang}}, \bibinfo {author}
  {\bibfnamefont {H.}~\bibnamefont {Zhang}}, \bibinfo {author} {\bibfnamefont
  {Z.}~\bibnamefont {Xu}}, \bibinfo {author} {\bibfnamefont {J.}~\bibnamefont
  {Denlinger}}, \bibinfo {author} {\bibfnamefont {A.}~\bibnamefont {Fedorov}},
  \bibinfo {author} {\bibfnamefont {H.}~\bibnamefont {Yang}}, \bibinfo {author}
  {\bibfnamefont {W.}~\bibnamefont {Duan}}, \bibinfo {author} {\bibfnamefont
  {H.}~\bibnamefont {Yao}}, \bibinfo {author} {\bibfnamefont {Y.}~\bibnamefont
  {Wu}}, \bibinfo {author} {\bibfnamefont {y.~S.}\ \bibnamefont {Fan}},
  \bibinfo {author} {\bibfnamefont {H.}~\bibnamefont {Zhang}}, \bibinfo
  {author} {\bibfnamefont {X.}~\bibnamefont {Chen}}, \ and\ \bibinfo {author}
  {\bibfnamefont {S.}~\bibnamefont {Zhou}},\ }\href
  {http://arxiv.org/abs/1603.08508} {\ }\Eprint
  {http://arxiv.org/abs/1603.08508} {arXiv:1603.08508} \BibitemShut {NoStop}%
\bibitem [{\citenamefont {Liang}\ \emph {et~al.}()\citenamefont {Liang},
  \citenamefont {Huang}, \citenamefont {Nie}, \citenamefont {Ding},
  \citenamefont {Gao}, \citenamefont {Hu}, \citenamefont {He}, \citenamefont
  {Zhang}, \citenamefont {Wang}, \citenamefont {Shen}, \citenamefont {Liu},
  \citenamefont {Ai}, \citenamefont {Sun}, \citenamefont {Zhao}, \citenamefont
  {Lv}, \citenamefont {Liu}, \citenamefont {Li}, \citenamefont {Zhang},
  \citenamefont {Hu}, \citenamefont {Xu}, \citenamefont {Zhao}, \citenamefont
  {Liu}, \citenamefont {Mao}, \citenamefont {Jia}, \citenamefont {Zhang},
  \citenamefont {Yang}, \citenamefont {Wang}, \citenamefont {Peng},
  \citenamefont {Dai}, \citenamefont {Fang}, \citenamefont {Xu}, \citenamefont
  {Chen}, \citenamefont {Zhou}, \citenamefont {Physics},\ and\ \citenamefont
  {Orleans}}]{Liang2016}%
  \BibitemOpen
  \bibfield  {author} {\bibinfo {author} {\bibfnamefont {A.}~\bibnamefont
  {Liang}}, \bibinfo {author} {\bibfnamefont {J.}~\bibnamefont {Huang}},
  \bibinfo {author} {\bibfnamefont {S.}~\bibnamefont {Nie}}, \bibinfo {author}
  {\bibfnamefont {Y.}~\bibnamefont {Ding}}, \bibinfo {author} {\bibfnamefont
  {Q.}~\bibnamefont {Gao}}, \bibinfo {author} {\bibfnamefont {C.}~\bibnamefont
  {Hu}}, \bibinfo {author} {\bibfnamefont {S.}~\bibnamefont {He}}, \bibinfo
  {author} {\bibfnamefont {Y.}~\bibnamefont {Zhang}}, \bibinfo {author}
  {\bibfnamefont {C.}~\bibnamefont {Wang}}, \bibinfo {author} {\bibfnamefont
  {B.}~\bibnamefont {Shen}}, \bibinfo {author} {\bibfnamefont {J.}~\bibnamefont
  {Liu}}, \bibinfo {author} {\bibfnamefont {P.}~\bibnamefont {Ai}}, \bibinfo
  {author} {\bibfnamefont {X.}~\bibnamefont {Sun}}, \bibinfo {author}
  {\bibfnamefont {W.}~\bibnamefont {Zhao}}, \bibinfo {author} {\bibfnamefont
  {S.}~\bibnamefont {Lv}}, \bibinfo {author} {\bibfnamefont {D.}~\bibnamefont
  {Liu}}, \bibinfo {author} {\bibfnamefont {C.}~\bibnamefont {Li}}, \bibinfo
  {author} {\bibfnamefont {Y.}~\bibnamefont {Zhang}}, \bibinfo {author}
  {\bibfnamefont {Y.}~\bibnamefont {Hu}}, \bibinfo {author} {\bibfnamefont
  {Y.}~\bibnamefont {Xu}}, \bibinfo {author} {\bibfnamefont {L.}~\bibnamefont
  {Zhao}}, \bibinfo {author} {\bibfnamefont {G.}~\bibnamefont {Liu}}, \bibinfo
  {author} {\bibfnamefont {Z.}~\bibnamefont {Mao}}, \bibinfo {author}
  {\bibfnamefont {X.}~\bibnamefont {Jia}}, \bibinfo {author} {\bibfnamefont
  {S.}~\bibnamefont {Zhang}}, \bibinfo {author} {\bibfnamefont
  {F.}~\bibnamefont {Yang}}, \bibinfo {author} {\bibfnamefont {Z.}~\bibnamefont
  {Wang}}, \bibinfo {author} {\bibfnamefont {Q.}~\bibnamefont {Peng}}, \bibinfo
  {author} {\bibfnamefont {X.}~\bibnamefont {Dai}}, \bibinfo {author}
  {\bibfnamefont {Z.}~\bibnamefont {Fang}}, \bibinfo {author} {\bibfnamefont
  {Z.}~\bibnamefont {Xu}}, \bibinfo {author} {\bibfnamefont {C.}~\bibnamefont
  {Chen}}, \bibinfo {author} {\bibfnamefont {X.~J.}\ \bibnamefont {Zhou}},
  \bibinfo {author} {\bibfnamefont {E.}~\bibnamefont {Physics}}, \ and\
  \bibinfo {author} {\bibfnamefont {N.}~\bibnamefont {Orleans}},\ }\href@noop
  {} {\ }\Eprint {http://arxiv.org/abs/arXiv:1604.01706v1}
  {arXiv:1604.01706} \BibitemShut {NoStop}%
\bibitem [{\citenamefont {Ali}\ \emph {et~al.}(2014)\citenamefont {Ali},
  \citenamefont {Xiong}, \citenamefont {Flynn}, \citenamefont {Tao},
  \citenamefont {Gibson}, \citenamefont {Schoop}, \citenamefont {Liang},
  \citenamefont {Haldolaarachchige}, \citenamefont {Hirschberger},
  \citenamefont {Ong},\ and\ \citenamefont {Cava}}]{Ali2014}%
  \BibitemOpen
  \bibfield  {author} {\bibinfo {author} {\bibfnamefont {M.~N.}\ \bibnamefont
  {Ali}}, \bibinfo {author} {\bibfnamefont {J.}~\bibnamefont {Xiong}}, \bibinfo
  {author} {\bibfnamefont {S.}~\bibnamefont {Flynn}}, \bibinfo {author}
  {\bibfnamefont {J.}~\bibnamefont {Tao}}, \bibinfo {author} {\bibfnamefont
  {Q.~D.}\ \bibnamefont {Gibson}}, \bibinfo {author} {\bibfnamefont {L.~M.}\
  \bibnamefont {Schoop}}, \bibinfo {author} {\bibfnamefont {T.}~\bibnamefont
  {Liang}}, \bibinfo {author} {\bibfnamefont {N.}~\bibnamefont
  {Haldolaarachchige}}, \bibinfo {author} {\bibfnamefont {M.}~\bibnamefont
  {Hirschberger}}, \bibinfo {author} {\bibfnamefont {N.~P.}\ \bibnamefont
  {Ong}}, \ and\ \bibinfo {author} {\bibfnamefont {R.~J.}\ \bibnamefont
  {Cava}},\ }\href {\doibase 10.1038/nature13763} {\bibfield  {journal}
  {\bibinfo  {journal} {Nature(London) }\ }\textbf {\bibinfo {volume} {514}},\ \bibinfo
  {pages} {205} (\bibinfo {year} {2014})}\BibitemShut {NoStop}%
\bibitem [{\citenamefont {Soluyanov}\ \emph {et~al.}(2015)\citenamefont
  {Soluyanov}, \citenamefont {Gresch}, \citenamefont {Wang}, \citenamefont
  {Wu}, \citenamefont {Troyer}, \citenamefont {Dai},\ and\ \citenamefont
  {Bernevig}}]{Soluyanov2015}%
  \BibitemOpen
  \bibfield  {author} {\bibinfo {author} {\bibfnamefont {A.~A.}\ \bibnamefont
  {Soluyanov}}, \bibinfo {author} {\bibfnamefont {D.}~\bibnamefont {Gresch}},
  \bibinfo {author} {\bibfnamefont {Z.}~\bibnamefont {Wang}}, \bibinfo {author}
  {\bibfnamefont {Q.}~\bibnamefont {Wu}}, \bibinfo {author} {\bibfnamefont
  {M.}~\bibnamefont {Troyer}}, \bibinfo {author} {\bibfnamefont
  {X.}~\bibnamefont {Dai}}, \ and\ \bibinfo {author} {\bibfnamefont {B.~A.}\
  \bibnamefont {Bernevig}},\ }\href {\doibase 10.1038/nature15768} {\bibfield
  {journal} {\bibinfo  {journal} {Nature(London)}\ }\textbf {\bibinfo {volume} {527}},\
  \bibinfo {pages} {495} (\bibinfo {year} {2015})}\BibitemShut {NoStop}%
\bibitem [{\citenamefont {Wu}\ \emph {et~al.}()\citenamefont {Wu},
  \citenamefont {Jo}, \citenamefont {Mou}, \citenamefont {Huang}, \citenamefont
  {Bud'ko}, \citenamefont {Canfield},\ and\ \citenamefont {Kaminski}}]{Wu2016}%
  \BibitemOpen
  \bibfield  {author} {\bibinfo {author} {\bibfnamefont {Y.}~\bibnamefont{Wu}}, 
  \bibinfo {author} {\bibfnamefont {D.}~\bibnamefont {Mou}} \bibinfo {author} {\bibfnamefont {N.~H.}\ \bibnamefont {Jo}}, \bibinfo
  {author} {\bibfnamefont {K.}~\bibnamefont {Sun} }, \bibinfo {author}
  {\bibfnamefont {L.}~\bibnamefont {Huang}}, \bibinfo {author} {\bibfnamefont
  {S.~L.}\ \bibnamefont {Bud'ko}}, \bibinfo {author} {\bibfnamefont {P.~C.}\
  \bibnamefont {Canfield}}, \ and\ \bibinfo {author} {\bibfnamefont
  {A.}~\bibnamefont {Kaminski}},\ }\href {http://arxiv.org/abs/1604.05176} {\
  }\Eprint {http://arxiv.org/abs/1604.05176} {Phys. Rev. B {\bf 94}, 121113 (2016)} \BibitemShut
  {NoStop}%
\bibitem [{\citenamefont {Xu}\ \emph {et~al.}(2015)\citenamefont {Xu},
  \citenamefont {Zhang},\ and\ \citenamefont {Zhang}}]{Xu2015a}%
  \BibitemOpen
  \bibfield  {author} {\bibinfo {author} {\bibfnamefont {Y.}~\bibnamefont
  {Xu}}, \bibinfo {author} {\bibfnamefont {F.}~\bibnamefont {Zhang}}, \ and\
  \bibinfo {author} {\bibfnamefont {C.}~\bibnamefont {Zhang}},\ }\href
  {\doibase 10.1103/PhysRevLett.115.265304} {\bibfield  {journal} {\bibinfo
  {journal} {Phys. Rev. Lett.}\ }\textbf {\bibinfo {volume} {115}},\ \bibinfo
  {pages} {265304} (\bibinfo {year} {2015})}\BibitemShut {NoStop}%
\bibitem [{\citenamefont {Yu}\ \emph {et~al.}()\citenamefont {Yu},
  \citenamefont {Yao},\ and\ \citenamefont {Yang}}]{Yu2016}%
  \BibitemOpen
  \bibfield  {author} {\bibinfo {author} {\bibfnamefont {Z.-M.}~\bibnamefont
  {Yu}}, \bibinfo {author} {\bibfnamefont {Y.}~\bibnamefont {Yao}}, \ and\
  \bibinfo {author} {\bibfnamefont {S.~A.}\ \bibnamefont {Yang}},\ }\href
  {http://arxiv.org/abs/1604.04030} {\ }\Eprint
  {http://arxiv.org/abs/1604.04030} {Phys. Rev. Lett. 117,077202} \BibitemShut {NoStop}%
\bibitem [{\citenamefont {Volovik}()}]{Volovik2016}%
  \BibitemOpen
  \bibfield  {author} {\bibinfo {author} {\bibfnamefont {G.~E.}\ \bibnamefont
  {Volovik}},\ }\href {http://arxiv.org/abs/1604.00849} {\ }\Eprint
  {http://arxiv.org/abs/1604.00849} {arXiv:1604.00849} \BibitemShut {NoStop}%
\bibitem [{\citenamefont {Udagawa}\ and\ \citenamefont
  {Bergholtz}()}]{Udagawa2016}%
  \BibitemOpen
  \bibfield  {author} {\bibinfo {author} {\bibfnamefont {M.}~\bibnamefont
  {Udagawa}}\ and\ \bibinfo {author} {\bibfnamefont {E.~J.}\ \bibnamefont
  {Bergholtz}},\ }\href {http://arxiv.org/abs/1604.08457} {\ }\Eprint
  {http://arxiv.org/abs/1604.08457} {Phys. Rev. Lett. {\bf 117}, 086401} \BibitemShut {NoStop}%
\bibitem [{\citenamefont {Zyuzin}\ and\ \citenamefont {Tiwari}()}]{Zyuzin2016}%
  \BibitemOpen
  \bibfield  {author} {\bibinfo {author} {\bibfnamefont {A.~A.}\ \bibnamefont
  {Zyuzin}}\ and\ \bibinfo {author} {\bibfnamefont {R.~P.}\ \bibnamefont
  {Tiwari}},\ }\href {http://arxiv.org/abs/1601.00890} {\ }\Eprint
  {http://arxiv.org/abs/1601.00890} {JETP Lett. 103, 717 (2016)} \BibitemShut {NoStop}%
\bibitem [{\citenamefont {Koshino}(2016)}]{Koshino2016}%
  \BibitemOpen
  \bibfield  {author} {\bibinfo {author} {\bibfnamefont {M.}~\bibnamefont
  {Koshino}},\ }\href {\doibase 10.1103/PhysRevB.94.035202} {\bibfield
  {journal} {\bibinfo  {journal} {Phys. Rev. B}\ }\textbf {\bibinfo {volume}
  {94}},\ \bibinfo {pages} {035202} (\bibinfo {year} {2016})}\BibitemShut
  {NoStop}%
\bibitem [{\citenamefont {Brien}\ \emph {et~al.}(2016)\citenamefont {Brien},
  \citenamefont {Diez},\ and\ \citenamefont {Beenakker}}]{Brien2016}%
  \BibitemOpen
  \bibfield  {author} {\bibinfo {author} {\bibfnamefont {T.~E.}\
  \bibnamefont {O'Brien}}, \bibinfo {author} {\bibfnamefont {M.}~\bibnamefont
  {Diez}}, \ and\ \bibinfo {author} {\bibfnamefont {C.~W.~J.}\ \bibnamefont
  {Beenakker}},\ }\href {\doibase 10.1103/PhysRevLett.116.236401} {\bibfield
  {journal} {\bibinfo  {journal} {Phys. Rev. Lett.}\ }\textbf {\bibinfo
  {volume} {116}},\ \bibinfo {pages} {236401} (\bibinfo {year}
  {2016})}\BibitemShut {NoStop}%
\bibitem [{\citenamefont {Levitov}\ \emph {et~al.}(1985)\citenamefont
  {Levitov}, \citenamefont {Nazarov},\ and\ \citenamefont
  {Eliashberg}}]{Levitov1985}%
  \BibitemOpen
  \bibfield  {author} {\bibinfo {author} {\bibfnamefont {L.}~\bibnamefont
  {Levitov}}, \bibinfo {author} {\bibfnamefont {Y.}~\bibnamefont {Nazarov}}, \
  and\ \bibinfo {author} {\bibfnamefont {G.}~\bibnamefont {Eliashberg}},\
  }\href
  {http://scholar.google.com/scholar?hl=en&btnG=Search&q=intitle:Magnetoelectric+effects+in+conductors+with+mirror+isomer+symmetry#0}
  {\bibfield  {journal} {\bibinfo  {journal} {Zh. Eksp. Teor. Fiz.}\ }\textbf
  {\bibinfo {volume} {88}},\ \bibinfo {pages} {229} (\bibinfo {year}
  {1985})}\BibitemShut {NoStop}%
\bibitem [{\citenamefont {Edelstein}(1990)}]{Edelstein1990}%
  \BibitemOpen
  \bibfield  {author} {\bibinfo {author} {\bibfnamefont {V.}~\bibnamefont
  {Edelstein}},\ }\href {\doibase 10.1016/0038-1098(90)90963-C} {\bibfield
  {journal} {\bibinfo  {journal} {Solid State Commun.}\ }\textbf {\bibinfo
  {volume} {73}},\ \bibinfo {pages} {233} (\bibinfo {year} {1990})}\BibitemShut
  {NoStop}%
\bibitem [{\citenamefont {Fujimoto}\ and\ \citenamefont
  {Yip}(2012)}]{Fujimoto2012}%
  \BibitemOpen
  \bibfield  {author} {\bibinfo {author} {\bibfnamefont {S.}~\bibnamefont
  {Fujimoto}}\ and\ \bibinfo {author} {\bibfnamefont {S.~K.}\ \bibnamefont
  {Yip}},\ }\href@noop {} {\emph {\bibinfo {title} {{{\rm Chapter 8 in} {\it
  Non-centrosymmetric Superconductors}}}}},\ edited by\ \bibinfo {editor}
  {\bibfnamefont {E.}\ \bibnamefont {Bauer}}\ and\ \bibinfo {editor}
  {\bibnamefont {M.~Sigrist}}\ (\bibinfo  {publisher} {Springer},\ \bibinfo
  {address} {Berlin},\ \bibinfo {year} {2012} )\BibitemShut {NoStop}%
\bibitem [{\citenamefont {Edelstein}(1989)}]{Edelstein1989}%
  \BibitemOpen
  \bibfield  {author} {\bibinfo {author} {\bibfnamefont {V.~M.}\ \bibnamefont
  {Edelstein}},\ }\href@noop {} {\bibfield  {journal} {\bibinfo  {journal}
  {Sov. Phys. JETP}\ }\textbf {\bibinfo {volume} {68}},\ \bibinfo {pages}
  {1244} (\bibinfo {year} {1989})}\BibitemShut {NoStop}%
\bibitem [{\citenamefont {Edelstein}(1995)}]{Edelstein1995}%
  \BibitemOpen
  \bibfield  {author} {\bibinfo {author} {\bibfnamefont {V.~M.}\ \bibnamefont
  {Edelstein}},\ }\href {\doibase 10.1103/PhysRevLett.75.2004} {\bibfield
  {journal} {\bibinfo  {journal} {Phys. Rev. Lett.}\ }\textbf {\bibinfo
  {volume} {75}},\ \bibinfo {pages} {2004} (\bibinfo {year}
  {1995})}\BibitemShut {NoStop}%
\bibitem [{\citenamefont {Yip}(2002)}]{Yip2002}%
  \BibitemOpen
  \bibfield  {author} {\bibinfo {author} {\bibfnamefont {S.~K.}\ \bibnamefont
  {Yip}},\ }\href {\doibase 10.1103/PhysRevB.65.144508} {\bibfield  {journal}
  {\bibinfo  {journal} {Phys. Rev. B}\ }\textbf {\bibinfo {volume} {65}},\
  \bibinfo {pages} {144508} (\bibinfo {year} {2002})}, \BibitemShut
  {NoStop}%
\bibitem [{\citenamefont {Yip}(2005)}]{Yip2005}%
  \BibitemOpen
  \bibfield  {author} {\bibinfo {author} {\bibfnamefont {S.~K.}\ \bibnamefont
  {Yip}},\ }\href {\doibase 10.1007/s10909-005-6012-7} {\bibfield  {journal}
  {\bibinfo  {journal} {J. Low Temp. Phys.}\ }\textbf {\bibinfo {volume}
  {140}},\ \bibinfo {pages} {67} (\bibinfo {year} {2005})}\BibitemShut
  {NoStop}%
\bibitem [{\citenamefont {Fujimoto}(2005)}]{Fujimoto2005}%
  \BibitemOpen
  \bibfield  {author} {\bibinfo {author} {\bibfnamefont {S.}~\bibnamefont
  {Fujimoto}},\ }\href {\doibase 10.1103/PhysRevB.72.024515} {\bibfield
  {journal} {\bibinfo  {journal} {Phys. Rev. B}\ }\textbf {\bibinfo {volume}
  {72}},\ \bibinfo {pages} {024515} (\bibinfo {year} {2005})}\BibitemShut
  {NoStop}%
\bibitem [{\citenamefont {Edelstein}(2005)}]{Edelstein2005}%
  \BibitemOpen
  \bibfield  {author} {\bibinfo {author} {\bibfnamefont {V.~M.}\ \bibnamefont
  {Edelstein}},\ }\href {\doibase 10.1103/PhysRevB.72.172501} {\bibfield
  {journal} {\bibinfo  {journal} {Phys. Rev. B}\ }\textbf {\bibinfo {volume}
  {72}},\ \bibinfo {pages} {172501} (\bibinfo {year} {2005})}\BibitemShut
  {NoStop}%
\bibitem [{\citenamefont {Fujimoto}(2007)}]{Fujimoto2007}%
  \BibitemOpen
  \bibfield  {author} {\bibinfo {author} {\bibfnamefont {S.}~\bibnamefont
  {Fujimoto}},\ }\href {\doibase 10.1143/JPSJ.76.034712} {\bibfield  {journal}
  {\bibinfo  {journal} {J. Phys. Soc. Japan}\ }\textbf {\bibinfo {volume}
  {76}},\ \bibinfo {pages} {034712} (\bibinfo {year} {2007})}\BibitemShut
  {NoStop}%
\bibitem [{\citenamefont {Maki}(1963)}]{Maki1963}%
  \BibitemOpen
  \bibfield  {author} {\bibinfo {author} {\bibfnamefont {K.}~\bibnamefont
  {Maki}},\ }\href {\doibase 10.1143/PTP.29.10} {\bibfield  {journal} {\bibinfo
   {journal} {Prog. Theor. Phys.}\ }\textbf {\bibinfo {volume} {29}},\ \bibinfo
  {pages} {10} (\bibinfo {year} {1963})}\BibitemShut {NoStop}%
\bibitem [{\citenamefont {Tretiakov}\ \emph {et~al.}(2008)\citenamefont
  {Tretiakov}, \citenamefont {Clarke}, \citenamefont {Chern}, \citenamefont
  {Bazaliy},\ and\ \citenamefont {Tchernyshyov}}]{Tretiakov2008}%
  \BibitemOpen
  \bibfield  {author} {\bibinfo {author} {\bibfnamefont {O.~A.}\ \bibnamefont
  {Tretiakov}}, \bibinfo {author} {\bibfnamefont {D.}~\bibnamefont {Clarke}},
  \bibinfo {author} {\bibfnamefont {G.~W.}\ \bibnamefont {Chern}}, \bibinfo
  {author} {\bibfnamefont {Y.~B.}\ \bibnamefont {Bazaliy}}, \ and\ \bibinfo
  {author} {\bibfnamefont {O.}~\bibnamefont {Tchernyshyov}},\ }\href {\doibase
  10.1103/PhysRevLett.100.127204} {\bibfield  {journal} {\bibinfo  {journal}
  {Phys. Rev. Lett.}\ }\textbf {\bibinfo {volume} {100}},\ \bibinfo {pages}
  {127204} (\bibinfo {year} {2008})}\BibitemShut {NoStop}%
\bibitem [{\citenamefont {Everschor}\ \emph {et~al.}(2012)\citenamefont
  {Everschor}, \citenamefont {Garst}, \citenamefont {Binz}, \citenamefont
  {Jonietz}, \citenamefont {M{\"{u}}hlbauer}, \citenamefont {Pfleiderer},\ and\
  \citenamefont {Rosch}}]{Everschor2012}%
  \BibitemOpen
  \bibfield  {author} {\bibinfo {author} {\bibfnamefont {K.}~\bibnamefont
  {Everschor}}, \bibinfo {author} {\bibfnamefont {M.}~\bibnamefont {Garst}},
  \bibinfo {author} {\bibfnamefont {B.}~\bibnamefont {Binz}}, \bibinfo {author}
  {\bibfnamefont {F.}~\bibnamefont {Jonietz}}, \bibinfo {author} {\bibfnamefont
  {S.}~\bibnamefont {M{\"{u}}hlbauer}}, \bibinfo {author} {\bibfnamefont
  {C.}~\bibnamefont {Pfleiderer}}, \ and\ \bibinfo {author} {\bibfnamefont
  {A.}~\bibnamefont {Rosch}},\ }\href {\doibase 10.1103/PhysRevB.86.054432}
  {\bibfield  {journal} {\bibinfo  {journal} {Phys. Rev. B}\ }\textbf {\bibinfo
  {volume} {86}},\ \bibinfo {pages} {054432} (\bibinfo {year}
  {2012})}\BibitemShut {NoStop}%
\bibitem [{\citenamefont {Schulz}\ \emph {et~al.}(2012)\citenamefont {Schulz},
  \citenamefont {Ritz}, \citenamefont {Bauer}, \citenamefont {Halder},
  \citenamefont {Wagner}, \citenamefont {Franz}, \citenamefont {Pfleiderer},
  \citenamefont {Everschor}, \citenamefont {Garst},\ and\ \citenamefont
  {Rosch}}]{Schulz2012a}%
  \BibitemOpen
  \bibfield  {author} {\bibinfo {author} {\bibfnamefont {T.}~\bibnamefont
  {Schulz}}, \bibinfo {author} {\bibfnamefont {R.}~\bibnamefont {Ritz}},
  \bibinfo {author} {\bibfnamefont {A.}~\bibnamefont {Bauer}}, \bibinfo
  {author} {\bibfnamefont {M.}~\bibnamefont {Halder}}, \bibinfo {author}
  {\bibfnamefont {M.}~\bibnamefont {Wagner}}, \bibinfo {author} {\bibfnamefont
  {C.}~\bibnamefont {Franz}}, \bibinfo {author} {\bibfnamefont
  {C.}~\bibnamefont {Pfleiderer}}, \bibinfo {author} {\bibfnamefont
  {K.}~\bibnamefont {Everschor}}, \bibinfo {author} {\bibfnamefont
  {M.}~\bibnamefont {Garst}}, \ and\ \bibinfo {author} {\bibfnamefont
  {A.}~\bibnamefont {Rosch}},\ }\href {\doibase 10.1038/nphys2231} {\bibfield
  {journal} {\bibinfo  {journal} {Nat. Phys.}\ }\textbf {\bibinfo {volume}
  {8}},\ \bibinfo {pages} {301} (\bibinfo {year} {2012})}\BibitemShut {NoStop}%
\bibitem [{\citenamefont {Combescot}\ and\ \citenamefont
  {Dombre}(1986)}]{Combescot1986}%
  \BibitemOpen
  \bibfield  {author} {\bibinfo {author} {\bibfnamefont {R.}~\bibnamefont
  {Combescot}}\ and\ \bibinfo {author} {\bibfnamefont {T.}~\bibnamefont
  {Dombre}},\ }\href {\doibase 10.1103/PhysRevB.33.79} {\bibfield  {journal}
  {\bibinfo  {journal} {Phys. Rev. B}\ }\textbf {\bibinfo {volume} {33}},\
  \bibinfo {pages} {79} (\bibinfo {year} {1986})}\BibitemShut {NoStop}%
\end{thebibliography}

\end{document}